\documentclass[amsmath, amsthm, amssymb, floatfix, notitlepage, nofootinbib, twocolumn, 10pt]{revtex4-2}

\usepackage[utf8]{inputenc}

\newtheorem{lemma}{Lemma}

\newtheorem{definition}{Definition}
\newtheorem{conjecture}{Conjecture}
\newtheorem{proposition}{Proposition}

\usepackage{amsmath}
\usepackage{amssymb}
\usepackage{upgreek}
\usepackage{braket}
\usepackage{color}
\usepackage{graphicx}
\usepackage{tikz}
\usepackage{multirow}
\usetikzlibrary{quantikz}
\usepackage{hyperref}
\hypersetup{colorlinks,allcolors=black,linktoc=all}

\newcommand{\edit}[1]{{\color{black} #1}}

\begin{document}
\title{Error propagation in NISQ devices for solving classical optimization problems}
\author{Guillermo González-García$^{1, 2}$}
\email{guillermo.gonzalez@mpq.mpg.de}
\author{Rahul Trivedi$^{1, 2}$}
\email{rahul.trivedi@mpq.mpg.de}
\author{J.~Ignacio Cirac$^{1, 2}$}
\address{$^1$Max-Planck-Institut für Quantenoptik, Hans-Kopfermann-Str.~1, 85748 Garching, Germany.\\
$^2$Munich Center for Quantum Science and Technology (MCQST), Schellingstr. 4, D-80799 Munich, Germany.}
\date{\today}

\begin{abstract}
    We propose a random circuit model that attempts to capture the behaviour of noisy-intermediate scale quantum devices when used for variationally solving classical optimization problems. Our model accounts for the propagation of arbitrary single qubit errors through the circuit. We find that even with a small noise rate, the quality of the obtained optima implies that a single-qubit error rate of $1 / nD$ (where $n$ is the number of qubits and $D$ is the circuit depth) is needed for the possibility of a quantum advantage. We estimate that this translates to an error rate lower than $10^{-6}$ using QAOA for classical optimization problems with 2D circuits.
\end{abstract}
\maketitle

\section{Introduction}

Significant advances in the capabilities of quantum information processing hardware have been made recently, with the achievement of an important milestone of having reached quantum advantage \cite{google2019quantumsupremacy,quantum-advantage-pan,quantum-advantage-pan-2}. In addition to developing technologies towards the final goal of a fault tolerant quantum computer, there is widespread interest in exploring the capabilities of the currently available noisy intermediate-scale quantum (NISQ) devices \cite{Preskill2018NISQ}. This has inspired a number of heuristic quantum algorithms for NISQ devices \cite{2022-NISQ-algorithms,farhi2000adiabatic,albash2018_adiabatic_qc,2020_sycamore_vqe,cerezo2021variational,harrigan2021QAOA_non_planar,McClean_VQE_2016,Jones_VQA_NISQ2019,peruzzo2014variational,Wecker_VQA_2015,amaro2022filtering,annealing2_kadowaki_1998,Endo_NISQ2020,annealing1994_finnila,annealing2_kadowaki_1998,liu2022prospects_NISQ,houck2012chip_simulation,ippoliti_qsimulation_2021}, but it remains unclear if they can provide a quantum advantage for practically interesting problems. 

One of the proposed applications for NISQ devices is to solve combinatorial optimization problems. Since this class of problems contains NP-hard instances, it is considered hard to solve on classical computers \cite{korte2011combinatorial}. Quantum circuits can explore larger state spaces (e.g. entangled states) compared to their classical counterparts and hence there is a possibility of quantum speedup in some instances of these problems. Several heuristic variational quantum algorithms have been proposed for solving optimization problems \cite{cerezo2021variational,harrigan2021QAOA_non_planar,McClean_VQE_2016,Jones_VQA_NISQ2019,peruzzo2014variational,2020_sycamore_vqe,Wecker_VQA_2015,amaro2022filtering}, most notably the quantum approximate optimization algorithm (QAOA)\cite{farhi2014qaoa}. While variational algorithms in general lack provable guarantees for quantum speedups, there is a growing body of literature that suggests its use for practically interesting problems that remain hard to solve on classical computers \cite{farhi2019QAOA_supremacy,farhi2021QAOA_SK,guerreschi2019qaoa}, as well as provides evidence for its experimental feasibility  \cite{Zhou2020_QAOA,2020Guido_QAOA_ions}.
However, a number of algorithmic challenges to these heuristics which stem from the limitations on the accessible circuit architectures, such as barren plateaus \cite{cerezo2021barrenplateaus}, expressibility of the ansatz \cite{expressibility1_2021,expressibility2_2021_nakaji}, or reachability deficits \cite{reachabilitydeficits_QAOA_2020}, have been identified.

The influence of noise in the quantum devices on variational algorithms is an important consideration for assessing their utility in the near term. Noise-induced barren plateaus in the optimization landscape of the variational ansatz have been identified \cite{wang2021noise}. Numerical modelling of the impact of noise has also been attempted \cite{Marshall_2020_QAOA_noise,Cheng-Xue-QAOA-noise}, but this analysis is limited to very small circuit sizes. A set of rigorous results on the impact of noise in the quality of the optima obtained have been recently provided using entropic arguments  \cite{stilck2021limitations, aharonov1996limitations,ben2013quantum_refrigerator}. By simply tracking the von Neumann entropy of the quantum state, in Refs.~\cite{aharonov1996limitations,ben2013quantum_refrigerator} it was argued that beyond circuit depths of $\Theta(\text{log}(n))$ with depolarizing noise,  the output state is very close to the maximally mixed state. Ref.~\cite{stilck2021limitations} improved on this result and showed that even beyond circuit depths of $\Theta(1)$, the quality of solution of a large class of classical optimization algorithm obtained from this circuit can be achieved with a classical algorithm.

However, while these analyses already provide provable limits on the quality of variational quantum algorithms that can inform current experiments, they likely underestimate the impact of noise in variational circuits used in practice. This underestimation arises from the fact that these bounds are applicable to \emph{any} quantum circuit, and consequently they also bound performance of circuits which do not create significant entanglement in the quantum state. Therefore, these bounds do not capture propagation of errors through the quantum circuit. Furthermore, these bounds are also loose for noise models, such as amplitude damping noise, which can possibly \emph{decrease} the entropy in the quantum circuit --- however, for a typical quantum circuit, it is expected that the presence of such noise would still degrade its performance.

In this paper, we propose a random circuit model that attempts to analyze the impact of noise in NISQ devices used to variationally solve classical optimization problems. Unlike previously proposed random circuit models, our model captures the fact that these circuits map product states for product states i.e.~ideally, a typical member of this circuit family starts with a product state, builds entanglement in between the qubits and then disentangles it to another (product) state which is the bit-string that solves the classical optimization problem. Our study here is thus different from recent works that have studied the impact of noise in Haar-random quantum circuits \cite{bouland2021noise,deshpande2022tight,dalzell2021_white_noise,boixo2018characterizing_supremacy} with geometric constraints, where a typical member of these circuit families has an anti-concentrated output state, which is in stark contrast to what is expected for variational quantum circuits  that ideally output a state close to the ground state.

We perform an average case analysis on this distribution of circuits, and find that noise propagates very rapidly through the circuit, severely limiting the performance of the circuits. We provide both numerical results and analytical scalings for three different architectures: 1D, 2D, and nonlocal. Our results indicate that to obtain a solution to a classical optimization problem within a fixed multiplicative error of its true solution with a constant rate of noise $p$, the circuit depth of a typical member of this distribution has to be smaller than $\max\left(O(p^{-1/2}),O(1/(pn))\right)$ in 1D and $\max\left(O(p^{-1/3}),O(1/(pn))\right)$ in 2D. Furthermore, it follows from our analysis that the impact of error is qualitatively similar for different noise models in contrast to Ref.~\cite{stilck2021limitations}, which relies on the noise-channel being primitive and unital. Moreover, our analysis technique can easily be extended to understanding impact of noise on variational quantum algorithms for solving quantum optimization problem (i.e.~finding the ground state of a quantum many-body Hamiltonian).

In addition to this average case analysis, we provide a concentration result for low depth local circuits which shows that this average is representative of a typical circuit. However, we note that this does leave open the question of whether a clever method of choosing the parameters of a variational quantum circuit could be devised (for e.g. with a closed loop optimization of the noisy quantum circuit), such that the resulting circuit avoids the predicted proliferation of noise.

\section{Error propagation model}\label{sec:markov_chain}
\begin{figure*}
    \centering
    \includegraphics[scale=0.6]{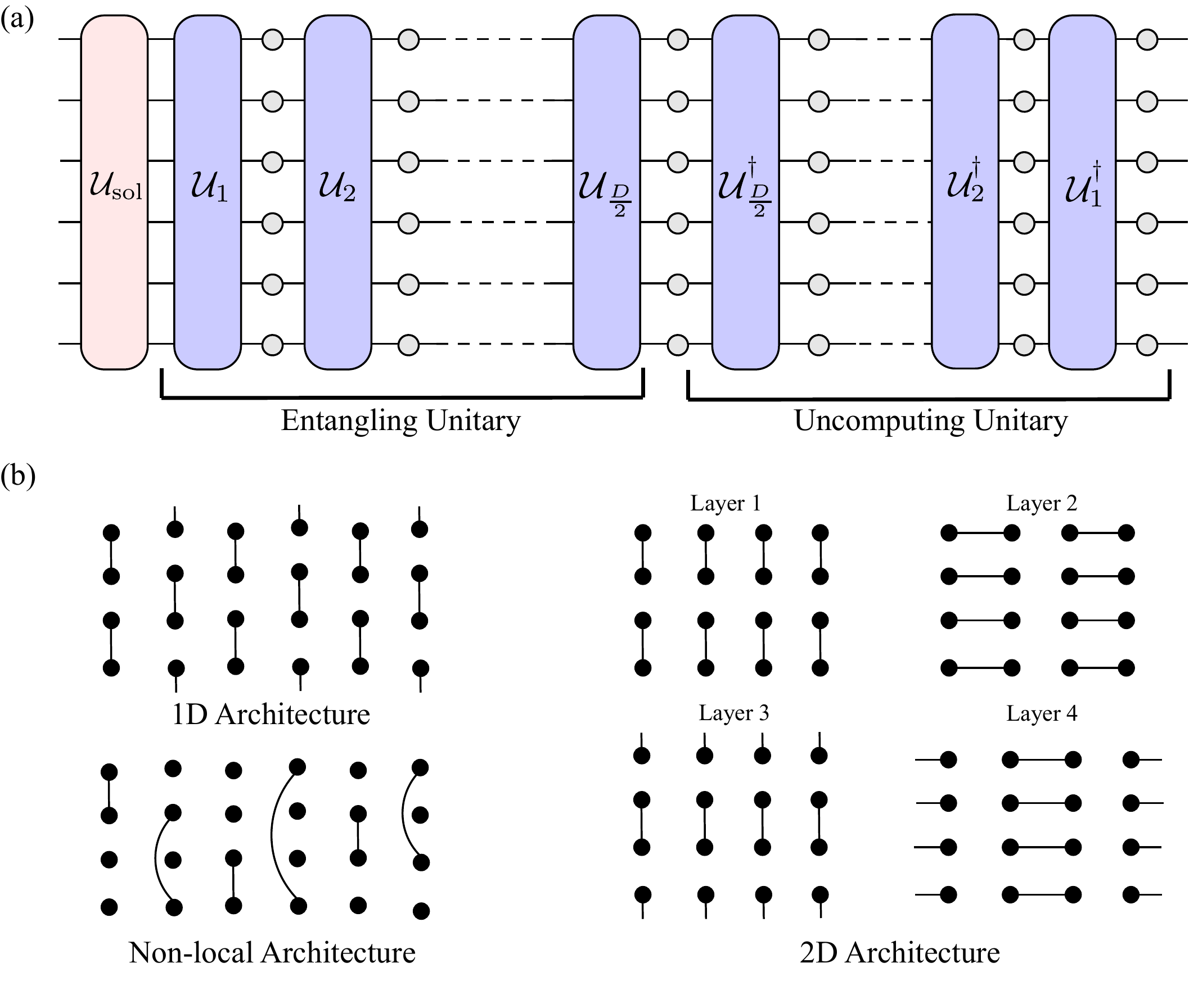}
    \caption{(a) Schematic depiction of the ensemble of unitary circuits considered in this paper. The circuits consist of an entangling and uncomputing unitary which maps a product state to another product state in the absence of noise. (b) Different circuit architectures (1D, 2D and non-local circuits) studied in this paper. All the circuit architectures are assumed to be translationally invariant --- in particular, 1D and 2D, we assume periodic boundary conditions)}
\label{fig:noise_model}
\end{figure*}

Consider the problem of solving a classical optimization problem over $\{0, 1\}^n$, and suppose that the unitary circuit which maps an initial product state $\ket{0}^{\otimes n}$ to the solution $x^* \in \{0, 1\}^n$ is given by $\mathcal{U}_\textnormal{sol}$. Since we are only interested in analyzing a lower bound on the impact of errors, we assume that the optimal circuit is already known and is noiseless, and do not address the problem of finding it. For depth $D$, we now consider a random quantum channel as $\Phi=\bigcirc_{t=1}^{D}T_{t}$, where the $T_{t}$ are given by

\begin{align}
T_{t}=\begin{cases}
        \mathcal{N}\circ\mathcal{U}^{(t)} \emph{ if } t \leq D/2,\\
       \mathcal{N}\circ \left[\mathcal{U}^{(t-D/2)}\right]^{\dagger} \emph{ if } t>D/2, \\
    \end{cases}
\end{align}
where $\mathcal{N}$ is a layer of the noise channel, and $\mathcal{U}^{(t)}$ is a layer of 2-qubit Haar random unitaries. The effective channel for the circuit under consideration, depicted in Fig.~\ref{fig:noise_model}a, will be constructed via $\Phi\circ \mathcal{U}_\text{sol}$ .

Each member of the ensemble of circuits defined above solves the optimization problem in the absence of noise (i.e.~when $\mathcal{N} = \textnormal{id}$). Furthermore, in the absence of noise, the circuit builds entanglement in the input state (which is a product state) for the first $D/2$ layers (which we call the \emph{entangling unitary}), after which it uncomputes these unitaries to finally obtain a product state. However, any error in the unitary circuit propagates and alters the final result. Since we are interested in the propagation of noise, we assume that the 1-qubit gates are noiseless.
We point out that a similar model can be constructed for quantum optimization problems, i.e.~finding the ground state of a quantum many-body hamiltonian, with $\mathcal{U}_\text{sol}$ being the circuit which maps $\ket{0}^{\otimes n}$ to the ground state. Even while assuming $\mathcal{U}_\text{sol}$ to be noiseless, the noisy entangling and unentangling unitaries will still capture propagation of errors within this model.

We will consider three different architectures for generating $\mathcal{U}^{(t)}$ which model different interaction ranges that can be accessed on physical hardware (Fig.~\ref{fig:noise_model}b):

\begin{itemize}
  \item A 1D local architecture with periodic boundary conditions, where alternating layers of nearest neighbour gates are applied.
  \item A 2D local architecture with periodic boundary conditions on a square lattice. Alternating layers of horizontal and vertical 2-qubit gates are applied.
  \item A nonlocal architecture where, for each layer, $n/2$ pairs are chosen at random, and $n/2$ 2-qubit gates are applied between the pairs.
\end{itemize}

Given that, in the absence of noise, all instances of the random unitaries produce the same unitary, we will consider the output of the channel averaged over the random unitaries, for each of the architectures,

\begin{align}
    \Phi_\text{avg}^{\mathcal{A}}\left(\rho\right)=\int_{\mathcal{U}}d\mathcal{U}\Phi\left(\rho\right). \ \text{with } \mathcal{A} \in \{\text{1D}, \text{2D}, \text{NL}\},
\end{align}
where $\Phi_{\mathrm{avg}}^{\mathrm{1D}}$ will represent the averaged channel with the 1D architecture, $\Phi_{\mathrm{avg}}^{\mathrm{2D}}$ the 2D architecture, and $\Phi_{\mathrm{avg}}^{\mathrm{NL}}$ the nonlocal architecture (Fig.~\ref{fig:noise_model}b).

%Here we analyze the output of the noise model defined above. We find that the output is simple: every qubit is either in its correct value, or completely depolarized (hence, in the maximally mixed state $I/2$). We will study  the number of qubits that are depolarized for a given system size $n$, circuit depth $D$, and error rate $p$, by mapping the problem to a Markov chain.

\begin{figure}
\centering
        \includegraphics[scale=0.6]{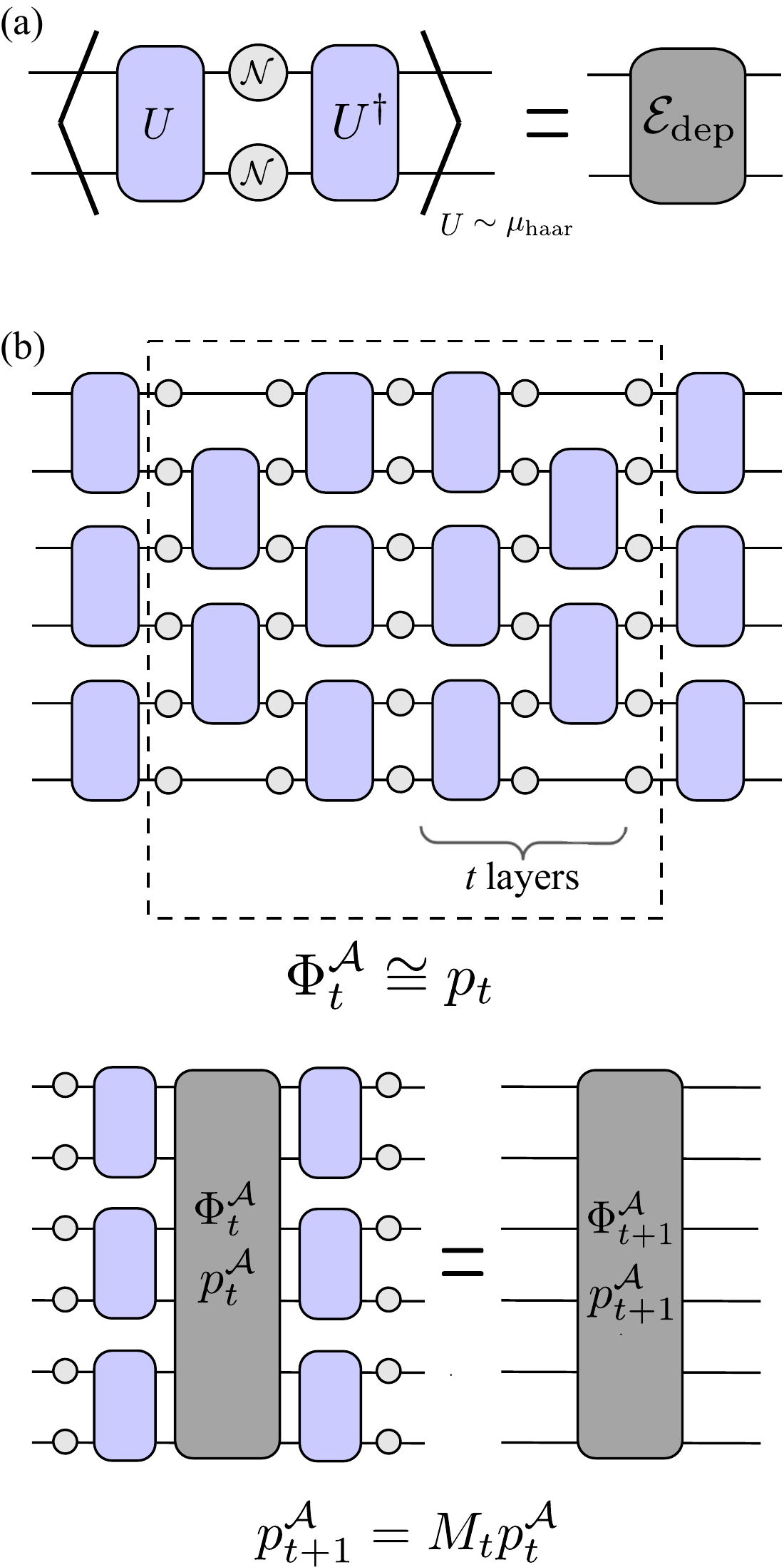}
        \caption{(a) Averaging over Haar-random unitary $U$ yields a 2-qubit depolarizing channel $\mathcal{E}_\textnormal{dep}$.
        (b) Schematic depiction of averaging over the circuit ensemble shown in Fig.~\ref{fig:noise_model} --- we average from the center of the circuit to outwards, and derive an effective Markov chain the describes the channels obtained.}
        \label{fig:building_block}
\end{figure}

An important fact that we will use to calculate $\Phi_\text{avg}^\mathcal{A}$ is that the twirl of a 2-qubit quantum channel $\mathcal{M}$ over the Haar measure is a 2-qubit depolarizing channel (Fig.~\ref{fig:building_block}a) \cite{Emerson_2005_channel_averaging} i.e.,
 %\begin{align}
 %\label{eqn:dep_channel}
 %\mathcal{E}_\textnormal{dep}(\rho) &= \int_{U\left(4\right)}\mu_\textnormal{Haar}\left(dU\right)U^{\dagger}\mathcal{N}_1 \otimes \mathcal{N}_2 \left(U\rho U^{\dagger}\right)U,\nonumber \\
 %& =\lambda\rho+\left(1-\lambda\right)\frac{I}{4},
 %\end{align}

 \begin{align}
 \label{eqn:twirl_channel}
 \mathcal{E}_\textnormal{dep} (X)&=\int_\mathcal{U} d\mathcal{U} \left[ \mathcal{U}^{\dagger} \mathcal{M} \mathcal{U}\right](X) \nonumber \\
 & =\lambda X+\left(1-\lambda\right)\frac{I^{\otimes 2}}{4} \text{tr}(X),
 \end{align}
\color{black}
where, if $A_k$ represents the $k$-th Kraus operator of the channel $\mathcal{M}$, then
 \begin{align}
\label{eqn:noise_strength}
     \lambda=\frac{1}{15}\left(\sum_{k}\left|\operatorname{tr}\left(A_{k}\right)\right|^{2}-1\right).
 \end{align}
Since this expression applies to arbitrary noise channels, it follows that the specific choice of noise does not greatly impact the qualitative behaviour of our model, with only the noise strength $\lambda$ depending on the specific channel.

In the remainder of this paper, we will consider that the noise is a depolarizing channel of strength $p$, 
\[
\mathcal{N}(\rho) = \left(1-p\right)\rho+p\  \textnormal{tr}(\rho) I/2,
\]
i.e. an error, interpreted at tracing out the qubit and replacing it with $I/2$, occurring with probability $p$. When this noise channel acts on two qubits, there are three distinct possibilities --- no errors occurred (i.e.~$\rho \to \rho$), both the qubits experienced errors ($\rho \to I^{\otimes 2}/4$) or only one qubit experienced error (i.e.~$\rho \to I/2 \otimes \text{tr}_1(\rho)$ or $ \text{tr}_2(\rho) \otimes I / 2$). Then, by direct application of Eqs.~(\ref{eqn:twirl_channel}) and (\ref{eqn:noise_strength}), we can compute the effective two-qubit depolarizing obtained after averaging over the two-qubit unitaries for these three cases :

\begin{itemize}
    \item For $\mathcal{M}(\rho) = \rho$ i.e.~no errors occur in either qubit between the application of $\mathcal{U}$ and $\mathcal{U}^{\dagger}$,  
    \begin{subequations}\label{eq:two_qubit_twirl_explicit}
    \begin{align}
    \mathcal{E}_\textnormal{dep}(\rho) = \rho.
    \end{align}
    \item If $\mathcal{M}(\rho) = \text{tr}(\rho) I^{\otimes 2}/4$ i.e.~errors occurred in both qubits between the application of $\mathcal{U}$ and $\mathcal{U}^{\dagger}$, 
    \begin{align}
    \mathcal{E}_\text{dep}(\rho) = \text{tr}(\rho) \frac{I^{\otimes 2}}{4}.
    \end{align}
    \item If $\mathcal{M}(\rho) = I / 2 \otimes \text{tr}_1(\rho) $ or $\text{tr}_2(\rho) \otimes I / 2$ i.e.~an error occurred in only in one of the qubits in between the application of $\mathcal{U}$ and $\mathcal{U}^\dagger$. Then, after averaging over $\mathcal{U}$ in Eq.~(\ref{eqn:twirl_channel}),
    \begin{align}
    \mathcal{E}_\text{dep}(\rho) = \frac{1}{5} \rho + \frac{4}{5} \text{tr}(\rho) \frac{I^{\otimes 2}}{4}.
    \end{align}
    \end{subequations}
    Physically, this can be interpreted as the error in one of the qubits propagating to the other qubit with probability $4/5$ due to the random  two-qubit entangling gates.
\end{itemize}

\color{black}

Let us now consider the quantum channel $\Phi_{\mathrm{avg}}^\mathcal{A}$ --- the key idea to computing this average channel, depicted schematically in Fig.~\ref{fig:building_block}b, is to start analyzing the circuit from its center (i.e.~in between the entangling and uncomputing unitary layers), and iteratively construct the channel obtained on including additional unitary layers. We denote by $\Phi^\mathcal{A}_t$ the channel formed by including and averaging over $t$ layers of the entangling and uncomputing unitary from the center of the circuit. For $t = 0$, we have only a single layer of depolarizing noise and hence $\Phi^{\mathcal{A}}_{t = 0} = \mathcal{N}^{\otimes n}$ --- this channel applies the noise independently on each qubit or leave their state unchanged. After including unitary layers and averaging over them, it follows from the analysis of the two qubit errors shown above that the resulting channels still either apply noise on each qubit or leaves them unchanged, although the noise is no longer applied independently on each qubit. Therefore, we assume the following ansatz for $\Phi_t^\mathcal{A}$
\begin{align}
\label{eqn:output_model}
\Phi_t^\mathcal{A} = \sum_{\vec{j}\in\{0, 1\}^n}p_t^\mathcal{A}(\vec{j}) \bigg[\bigotimes_{\alpha \in \vec{j} } \uptau_{\alpha}\bigg]
\end{align}
where
\[
\uptau_0(\cdot) = \text{id}(\cdot) \text{ and } \uptau_1(\cdot) = \text{tr}(\cdot) \frac{I}{2}.
\]
Here, $p^{\mathcal{A}}_t(\vec{j})$, for a given $\vec{j} \in \{0, 1\}^n$, is the probability that first qubit experiences noise if $j_0 = 1$ else remains unchanged, the second qubit experiences noise if $j_1 = 1$ else remains unchanged and so on. We note that at $t = 0$ i.e.~when $\Phi^{\mathcal{A}}_{t = 0}$ is a layer of depolarizing noise, 
\begin{align}\label{eq:initial_distr}
p_{t = 0}^\mathcal{A}(\vec{j}) = (1 - p)^{n - |\vec{j}|} p^{|\vec{j}|},
\end{align}
where $|\vec{j}| = \sum_{i = 1}^{n} j_i$.

Next, we show that if $\Phi^A_t$ is of the form of Eq.~(\ref{eqn:output_model}), then so is $\Phi^A_{t + 1}$ and relate the probability distribution $p_{t + 1}^{\mathcal{A}}$ to $p_t^{\mathcal{A}}$. Note that
\begin{align}\label{eq:next_step_from_prev}
\Phi^{\mathcal{A}}_{t + 1} = \mathbb{E}_{\mathcal{U}^{(t)}} \big[\mathcal{N}^{\otimes n} \mathcal{U}^{(t)\dagger} \Phi^{\mathcal{A}}_t  \mathcal{U}^{(t)}\mathcal{N}^{\otimes n}\big],
\end{align}
where $\mathcal{U}^{(t)}$ is the unitary applied at $t^{\textnormal{th}}$ layer from the center. Depending on the circuit architecture, $\mathcal{U}^{(t)} = \bigotimes_{(\alpha, \beta) \in S^{(t)}} \mathcal{U}^{(t)}_{\alpha, \beta}$, where $\mathcal{U}^{(t)}_{\alpha, \beta}$ is a two-qubit gate acting on qubits $\alpha, \beta$ and $S^{(t)}$ contains a list of qubits which interact with each other through the two-qubit unitaries. The averaging over each two-qubit unitary can now be done independently using Eq.~(\ref{eq:two_qubit_twirl_explicit}). In particular, consider averaging over one of these unitaries acting on qubits $\alpha$ and $\beta$
\begin{align*}
 &\mathbb{E}_{\mathcal{U}_{\alpha, \beta}^{(t)}}\big[ \mathcal{U}^{(t) \dagger}_{\alpha, \beta} \Phi^{\mathcal{A}}_t \mathcal{U}_{\alpha, \beta}^{(t)}\big] =\nonumber \\ &\sum_{\vec{j}}p^\mathcal{A}_t(\vec{j}) \bigg(\bigotimes_{\gamma \neq \alpha, \beta} \tau_\gamma \bigg) \otimes \mathbb{E}_{\mathcal{U}^{(t)}_{\alpha, \beta}}\big[ \mathcal{U}_{\alpha, \beta}^{(t)\dagger} \uptau_{j_\alpha} \otimes \uptau_{j_\beta} \mathcal{U}_{\alpha, \beta}^{(t)} \big]
\end{align*}
From Eq.~(\ref{eq:two_qubit_twirl_explicit}), it follows that
\begin{align*}
&\mathbb{E}_{\mathcal{U}^{(t)}_{\alpha, \beta}}\big[ \mathcal{U}_{\alpha, \beta}^{(t)\dagger} \uptau_{j_\alpha} \otimes \uptau_{j_\beta} \mathcal{U}_{\alpha, \beta}^{(t)}\big] \nonumber \\
&\qquad=\begin{cases}
\tau_{j_\alpha} \otimes \tau_{j_\alpha} & \text{if } j_\alpha = j_\beta, \\
\frac{1}{5}\tau_0 \otimes \tau_0 + \frac{4}{5}\tau_1 \otimes \tau_1 & \text{if } j_\alpha \neq j_\beta,
\end{cases}
\end{align*}
and therefore, it follows that
\[
\mathbb{E}_{\mathcal{U}_{\alpha, \beta}^{(t)}}\big[ \mathcal{U}^{(t) \dagger}_{\alpha, \beta} \Phi^{\mathcal{A}}_t \mathcal{U}_{\alpha, \beta}^{(t)}\big] = \sum_{\vec{j}}q_{\alpha, \beta; t}(\vec{j}) \bigg[\bigotimes_{\alpha \in \vec{j}} \uptau_\alpha \bigg],
\]
where the probability distribution $q_{\alpha, \beta; t}(\vec{j})$ is related to $p_t^\mathcal{A}(\vec{j})$ via a transition matrix that is identity on all but the $\alpha^\text{th}$ and $\beta^\text{th}$ bits, on which it is given by
\begin{align}\label{eq:transition_matrix_unitary}
&M^U_{\alpha, \beta}((j_\alpha, j_\beta) \to (j_\alpha', j_\beta')) \\ & \nonumber= \begin{cases}1 & \text{ if } (j_\alpha, j_\beta), (j_\alpha', j_\beta') = (0, 0),\\
1/5 &\text{ if } (j_\alpha, j_\beta) = (0, 1) \text{ or }(1, 0), (j_\alpha', j_\beta') = (0, 0),\\
4/5 &\text{ if } (j_\alpha, j_\beta) = (0, 1) \text{ or }(1, 0), (j_\alpha', j_\beta') = (1, 1), \\
1 & \text{ if }(j_\alpha, j_\beta), (j_\alpha', j_\beta') = (1, 1).
\end{cases}
\end{align}
Repeating this for all the two-qubit gates in the unitary $\mathcal{U}^{(t)}$, we obtain that
\[
\mathbb{E}_{\mathcal{U}^{(t)}} \big[\mathcal{U}^{(t)\dagger} \Phi^{\mathcal{A}}_t  \mathcal{U}^{(t)}\big] = \sum_{\vec{j}} q_t^\mathcal{A}(\vec{j}) \bigg[\bigotimes_{\alpha \in \vec{j}} \uptau_\alpha \bigg],
\]
where $q^\mathcal{A}_t(\vec{j})$ is a probability distribution related to $p^\mathcal{A}_t(\vec{j})$ via the transition matrix $M^U_t$ given by
\[
M^U_t = \bigotimes_{(\alpha, \beta) \in S^{(t)}} M^U_{\alpha, \beta}.
\]

Having averaged over the unitaries, we now consider applying the layers of depolarizing noise. This can be done explicitly, we obtain from Eq.~(\ref{eq:next_step_from_prev}) that
\[
\Phi^{\mathcal{A}}_{t + 1} = \sum_{\vec{j} \in \{0, 1\}^n} q_t^\mathcal{A}(\vec{j}) \bigg[ \bigotimes_{\alpha \in \vec{j}} \big(\mathcal{N} \uptau_\alpha \mathcal{N}\big)\bigg].
\]
Furthermore, it can be immediately seen from the definitions of $\uptau_\alpha$ that
\[
\mathcal{N} \uptau_\alpha \mathcal{N} = \begin{cases}
(1 - p)^2 \uptau_0 + (2p - p^2) \uptau_1 & \text{if } \alpha = 0, \\
\uptau_1 &\text{if } \alpha = 1.
\end{cases}
\]
Thus, the probability distribution $p^\mathcal{A}_{t + 1}(\vec{j})$ is obtained from the probability distribution $q^\mathcal{A}_{t}(\vec{j})$ using the transition matrix $M^\text{noise}$, which independently for every bit in $\vec{j}$ has the transition probabilities
\begin{align}\label{eq:transition_matrix_noise}
M^\text{noise}(j \to j') = \begin{cases} (1 - p)^2  & \text{if } j = 0, j' = 0, \\
2p - p^2 & \text{if } j = 0, j' = 1 \\
1 & \text{if } j = 1, j' = 1 \\
0 & \text{otherwise}.
\end{cases}
\end{align}
Combining the transition rules for both the unitary layer and the noise, we then obtain a Markov chain for the probability distribution $p^\mathcal{A}_t$,
\begin{align}\label{eqn:transition_matrix}
p^{\mathcal{A}}_{t + 1} = \big(M^\text{noise} M^U_t\big) p^{\mathcal{A}}_t.
\end{align}
To analyze the state obtained at the output of this circuit, after averaging over all the unitaries, we thus need only to compute the probabilities $p_{t = D/2}^\mathcal{A}$ starting from the initial distribution $p_{t = 0}^{\mathcal{A}}$ (Eq.~(\ref{eq:initial_distr})), which then gives us access to the channel $\Phi^{\mathcal{A}}_{t = D/2}$ (note that the effective time for which we need to evolve the Markov chain is half of the depth of the circuit, since each layer of Markov chain accounts for a layer in the entangling and uncomputing unitaries). This can be computed by simulating the Markov chain with transition matrices described above and allows us to analyze how many and which qubits in the solution of the optimization problem are, on average, correct. In the next section, we analyze this Markov chain both analytically and numerically to understand the impact of the noise and circuit depth on the quality of the output.

\section{Impact of errors}
\subsection{Analysis of error propagation}
\begin{figure}
    \centering
    \includegraphics[scale=0.5]{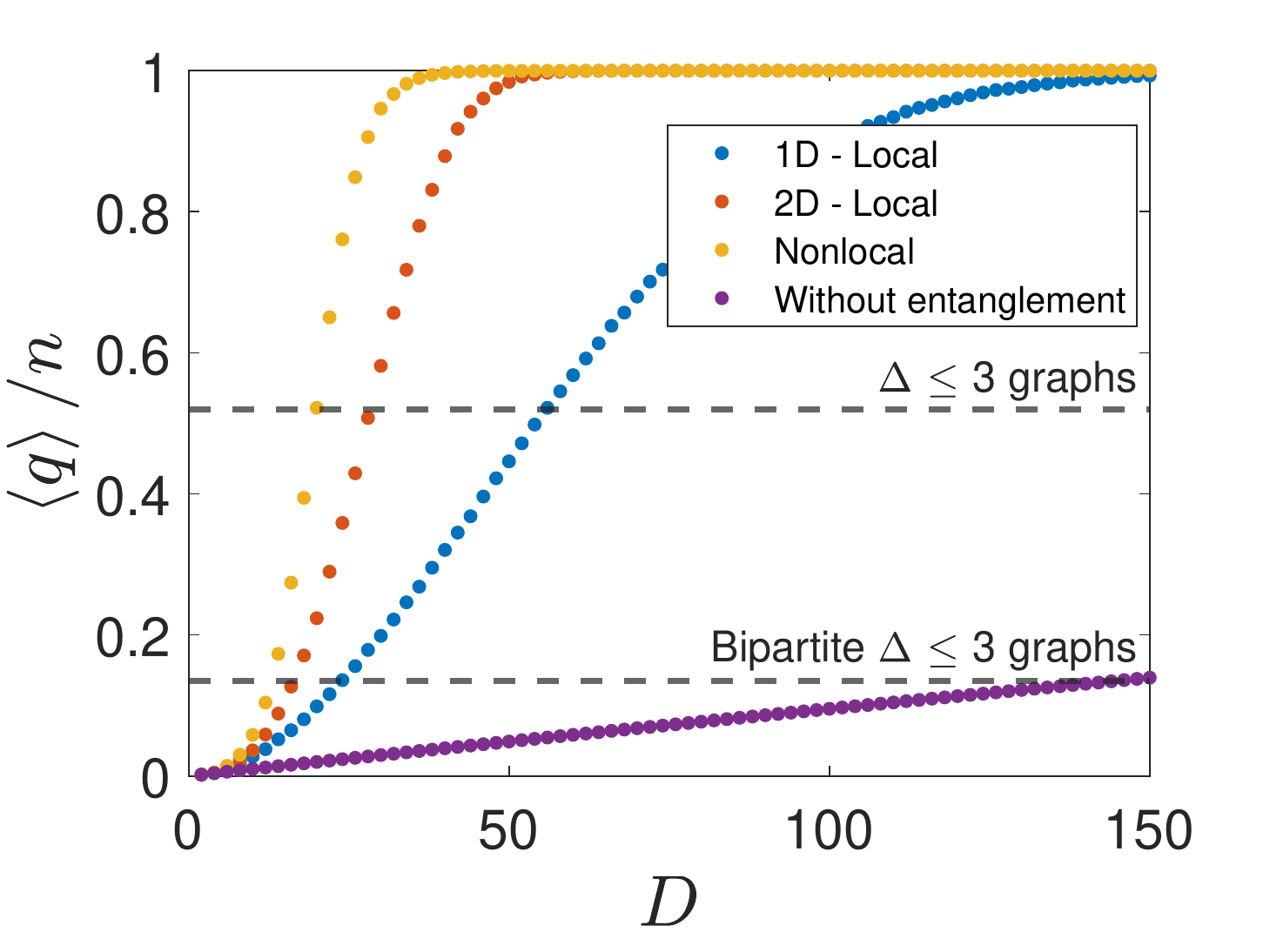}
    \caption{We represent the expectation value of the fraction of depolarized qubits, $\left<q\right>/n$, as a function of the circuit depth, for the different architectures: 1D, 2D, and nonlocal. We have used a system size $n=900$ and an error rate $p=10^{-3}$. Additionally, we have represented $\left<q\right>/n=1-\left(1-p\right)^D$, which is the number of depolarized qubits that one would get when applying only local depolarizing noise to all the qubits, without any unitaries (and therefore without entanglement). We see that the convergence to uniform is very fast with our model. The horizontal lines represent thresholds for classical superiority for unweighted Max-Cut problems on arbitrary bounded degree 3 graphs and bipartite bounded degree 3 graphs: for values of $\left<q\right>/n$ greater than the threshold, there is a classical efficient algorithm that outputs a better a better solution than the averaged quantum channel (see subsection \ref{subsection:implications_noisy_QAOA}). The number of samples taken is $2000$ which reduces the error in the estimated mean to 2\%.}
    \label{fig:1D-2D-nonlocal}
\end{figure}

We first compute the expectation value of the number of qubits that are depolarized at the end of the computation, which we denote as $\left<q\right>$. This can be done numerically by sampling from the Markov chain described in the previous section using Markov-chain Monte Carlo. The results are shown in Fig.~\ref{fig:1D-2D-nonlocal} --- our results show that, as expected, all the qubits depolarize exponentially fast with the circuit depths due to a rapid propagation of errors. 

\begin{table}[h!]
\centering
{\begin{tabular}{|c|c|c|c| }
 \hline
 & Architecture & Depth & $\left<q\right>/n$ \\
 \hline
  & 1D & $O(n)$ & $1-(1-p)^{O(D^2)}$ \\ 
 Shallow regime & 2D & $O(\sqrt{n})$ & $1-(1-p)^{O(D^3)}$ \\ 
 & NL & $O(\log n)$ & $1-(1-p)^{O(\exp(D))}$ \\ 
  \hline
  & 1D & $\Omega(n)$ & \multirow{3}{*}{$1-\left(1-p\right)^{O(nD)}$} \\ 
  Deep regime & 2D & $\Omega(\sqrt{n})$ & \\ 
  & NL & $\Omega(\log n)$  & \\ 
 \hline
\end{tabular}
\caption{Summary of the scaling of the expectation value of the fraction of depolarized qubits at the end of the computation, as a function of the circuit depth and system size. We identify two different regimes, a shallow depth regime and a deep regime. The deep regime represents the cases in which the light cone of a single error can reach the edges (i.e. a single error can propagate to all other qubits), while for the shallow regime circuits this is not the case.}
\label{table:scalings}}
\end{table}

In order to better understand these numerical results, we would like to analyze theoretically the behavior of the Markov chain. We first provide heuristic estimates for the scalings of $\langle q\rangle$ with $n$. Note that every time a qubit is depolarized in the effective Markov chain, the neighbouring qubit can be depolarized with a constant (i.e.~$\Theta(1)$) probability. Take, for instance, the 1D model --- at the end of the circuit, an error in the middle of the circuit will have propagated to $\min(O(D),n)$ qubits. We can now estimate the probability that a specific qubit will be depolarized at the end of the computation --- on average, it will be depolarized if at least one error has occurred in a qubit that is with a distance of $\min(O(D),n)$ of this qubit. Since there are $\Theta(D)$ steps of the Markov chain, the probability of a certain qubit not being depolarized is $\sim (1-p)^{O(D^2)}$ for circuits that are shallow, and $\sim (1-p)^{O(nD)}$ for deep circuits. Similary, in 2D a single error propagates on average to $\min(O(D^2),n)$ qubits, and in the nonlocal case to $\min(e^{O(D)},n)$. These scalings are displayed in Table \ref{table:scalings}. We note that, if the circuit is sufficiently deep, a single error can potentially propagate to all the other qubits, thus resulting in $\langle q \rangle$ scaling as $1-\left(1-p\right)^{O(nD)}$.

In addition to the scalings presented in Table \ref{table:scalings}, we provide a heuristic formula that works very well in practice for the 1D case:
\begin{align}
\label{eqn:heuristic_1D}
    \frac{\left<q\right>_{1D}}{n} \simeq \begin{cases} 
    1-(1-2p)^{\frac{9}{80} D^2} \text{ if }  D \leq \frac{5}{3}n\\
    1-(1-2p)^{\frac{3}{8}nD - \frac{5}{16}n^2} \text{ if }  D>\frac{5}{3}n.
    \end{cases}
\end{align}
 The derivation of this formula and its numerical verification is in appendix \ref{appendix:Heuristic_formula}. We also provide a semi-empirical formula for the 2D case, that has been obtained by fitting the data points to the expression in Table \ref{table:scalings}: 
\begin{align}
\label{eqn:empirical_2D}
    &\frac{\left<q\right>_{2D}}{n} \simeq \nonumber \\ 
    &\begin{cases} 
    1-\big(1-\frac{3}{2}p\big)^{0.026 D^3+0.054D^2} \text{ if }  D \leq 3.226\sqrt{n}\\
    1-\big(1-\frac{3}{2}p\big)^{\frac{1}{2}nD - 0.74n^{3/2}+0.56n} \text{ if }  D>3.226\sqrt{n}.
    \end{cases}
\end{align}
Finally, in appendix \ref{appendix:1-D_bound} we rigorously prove a lower bound on $\langle q\rangle$ for the 1D model with $D < n $ which has the same scaling with $p, D$ and $n$ as Eq.~(\ref{eqn:heuristic_1D}).

\subsection{Implications on circuit depths for noisy QAOA}
\label{subsection:implications_noisy_QAOA}

We apply our error model to the specific case of quantum circuits that try to use QAOA to solve classical optimization problems, and analyze how the propagation of errors could limit the performance of the algorithm. We will be considering the Max-Cut problem since it is  practically useful and hard to solve classically. Given a graph $G=\left(V,E\right)$ on $n$ vertices and adjacency matrix $a_{ij}$, finding the Max-Cut of $G$ is dividing the qubits into two groups, $V'$ and $V\setminus V'$, such that the sum of weights of edges between these two groups is maximal. It can equivalently be defined as the the bitstring $Z \in \{-1, 1\}^n$ maximizing the cost function
\begin{align}
    C=\frac{1}{2} \sum_{(i,j) \in E} a_{ij}\left(1-Z_iZ_j\right),
\end{align}
with the vertices corresponding to $Z = -1$ being in one group, and those corresponding to $Z = 1$ being in the other.

Let us denote by $C_{\mathrm{max}}$ the solution of this problem, and denote the average value of the cost function when random guessing by $C_\text{avg}$ i.e
\begin{align}
    C_{\mathrm{avg}}=\mathrm{tr}\left(C\frac{I}{2^n}\right)=\frac{1}{2} \sum_{(i,j) \in E} a_{ij}.
\end{align}
$C_\text{avg}$ would be the cost function obtained by a circuit in which all the qubits are depolarized. Since the output of the error propagation channel contains some qubits that are depolarized, the cost function obtained from the resulting state will be between $C_\text{avg}$ and $C_\text{max}$. As shown in appendix \ref{appendix:energy}, under the assumption that $a_{ij} > 0$, we are able to upper bound the average energy of the output as a function of the average number of depolarized qubits, $\left<q\right>$ as
\begin{widetext}
\begin{align}
    \label{eqn:energy_bound}
    \mathrm{tr}\left(C\Phi_{\mathrm{avg}}\left(\rho\right)\right) \leq \frac{1}{2}\left(1-\frac{\left<q\right>}{n}\right)\left(2-\frac{\left<q\right>}{n}\right)C_{\mathrm{max}}+\left(1-\left(1-\frac{\left<q\right>}{n}\right)^2\right)C_{\mathrm{avg}}.
\end{align}
\end{widetext}

%Furthermore, in appendix \ref{appendix:conc_bound}, we show that for low depth circuits, $\text{tr}(C\Phi(\rho))$ for a randomly drawn circuit instance concentrates around this mean indicating that it is representative of a typical circuit instance.

This expression readily provides an upper bound on the quality of the solution that we can compute given an error rate and a circuit depth. Given a lower bound on the maximum value $C_\text{max}$ of the cost function, it also allows us to upper bound the approximation ratio ($\alpha=\textnormal{Tr}[C\Phi_\text{avg}^\mathcal{A}(\rho_0)]/C_{\mathrm{max}}$). This approximation ratio can often inform of the existence of an efficient classical algorithm which obtains a similar solution --- for instance, the Goemans-Williamson algorithm is a classical approximation algorithm that has a performance guarantee of $\alpha > 0.878$ \cite{Goemans_MAXCUT}. It is thus a reasonable assumption to make that near-term quantum circuits are only useful if the approximation ratio, in the presence of noise, is better than those achievable by classical algorithms.

As a specific example, we briefly study unweighted (i.e.~$a_{ij} = 1$) bounded degree $\Delta = 3$ graphs which cannot be solved trivially in general. For any bounded degree $\Delta = 3$ graph the Edwards-Erd{\"o}s inequality provides a lower bound for the Max-Cut in terms of the number of edges, $C_{\mathrm{max}} \geq 2/3 |E|$ \cite{erdos1965inequality}. Combined with Eq. (\ref{eqn:energy_bound}), this provides an upper bound for the approximation ratio, 

\begin{align}
\label{eqn:approximation_ratio}
    \alpha \leq 1-\left(\frac{\left<q\right>}{2n}\right)^2.
\end{align} 

For degree 3 graphs, there is a classical approximation algorithm that achieves an approximation ratio of 0.9326 \cite{cubic_graphs_2004}. We therefore obtain that only when $\left<q\right>/n \leq 0.52$ can the quantum algorithm possibly output a better average energy than the classical one. That is, as soon as approximately half of the qubits are depolarized, we can be sure that the average quality of the solution is worse than the quality of the solution of classical approximation algorithms in the worst case. We represent an instance of this in Fig. \ref{fig:noise-1D-2D}. We note that this bound is not tight in every case, since we are considering all possible bounded degree $\Delta=3$ graphs. For example, if we consider only the bipartite ones, $C_{\mathrm{max}}=|E|$ and $C_{\mathrm{avg}}=|E|/2$, which gives an approximation ratio that is bounded by $\alpha \leq 1-\left<q\right>/\left(2n\right)$, and therefore $\left<q\right>/n \leq 0.135$. Therefore, in this case, it is already possible to certify the classical advantage when only around 15\% of the qubits are depolarized.

Using the scalings from Table \ref{table:scalings}, we obtain that for shallow circuits (as defined in the table), after a depth $D=O\left(\sqrt{1/p}\right)$ for 1D and $D=O\left(\sqrt[3]{1/p}\right)$ for 2D we already have a situation where the quality of the solution is worse than with classical approximation algorithms. This is respectively quadratically and cubically worse with respect to $p$ than the scaling reported in \cite{stilck2021limitations}. This is a consequence of the rapid spreading of the errors.

The impact of errors in 2D (and even all-to-all) architectures is much higher than in 1D due to a more rapid propagation of errors. \edit{However, this does not necessarily imply that 1D circuits are better for near term quantum computation since our analysis thus far does not account for the increased connectivity of the quantum circuit.} In order to take that into account, in the next subsection we consider a specific case, QAOA for solving a non-local problem, i.e a graph with long-range vertices (a non-planar graph). This necessarily places us in the deep circuit regime (as defined in Table \ref{table:scalings}) in all cases.

We stress that this is an average case analysis, which may not apply to particular situations or specific circuits. However, in appendix \ref{appendix:conc_bound} we derive a concentration bound for shallow local circuits,
\begin{align}
    \operatorname{Pr}\left[\left|C-\mathrm{tr}\left(C\Phi_{\mathrm{avg}}\left(\rho\right)\right)\right| \geq \alpha |E|\right] \leq 2 e^{-O\left(\frac{\alpha^{2}|E|}{\Delta^2D^{2k}}\right)},
\end{align}
where $\Delta$ is the degree if the graph and $k$ is the dimension of the architecture. Thus, for shallow local circuits, the output of a typical circuit is exponentially close to the average.

\subsubsection*{Required error rate}

\begin{figure*}
    \centering
    \includegraphics[scale=1.4]{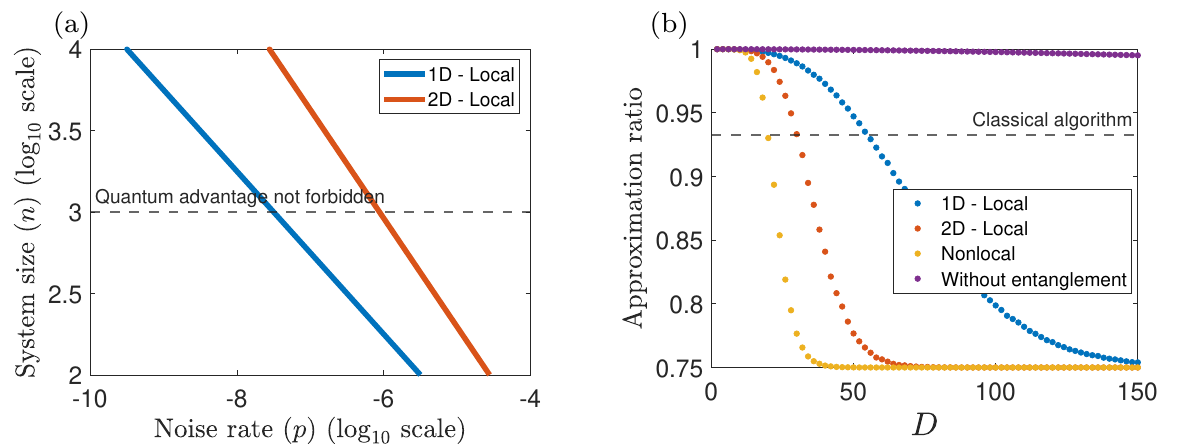}
    \caption{In (a) we represent the relation between the system size and the error rate when the average number of depolarized qubits at the end of the computation is fixed to $\left<q\right>/n=0.5$ for bounded degree $\Delta=3$ graphs. In the 1D case the depth is $D_{1D}=30n$, and in 2D $D_{2D}=10\sqrt{7n}$. The horizontal dashed line corresponds to system sizes of $n=1000$, which is when a potential quantum advantage could begin to be practically useful \cite{guerreschi2019qaoa}.
    In (b) we represent the upper bound for the approximation ratio of bounded degree $\Delta=3$ graphs, given by Eq. (\ref{eqn:approximation_ratio}), as a function of the circuit depth, for the different architectures: 1D, 2D, and nonlocal. We have used a system size $n=900$ and an error rate $p=10^{-3}$. The horizontal line represents the approximation ratio that is reachable by an efficient classical algorithm. The number of samples taken is $2000$ which reduces the error in the estimated mean to 2\%.}
    \label{fig:noise-1D-2D}
\end{figure*}

We also need to analyze the required circuit depth to run QAOA. Most quantum computing architectures have a planar design, similar to the 2D architecture that we are considering.  However, the optimization problems that are useful in practice consist usually of non-planar graphs. For example, planar Max-Cut can always be solved in polynomial time on classical computers \cite{1975_planar_maxcut}. Since non-planar graphs do not match the connectivity of the hardware, routing will be required to perform the computation. This can be done with SWAP gates that permute the different qubits, but comes at the cost of a growth in the circuit depth . Embedding a bounded degree graph on a square lattice results in an overhead of $\sqrt{n}$ in the gate count, while the cost is $n$ for the 1D local architecture \cite{harrigan2021QAOA_non_planar}. Furthermore, at least a few QAOA layers are necessary for the algorithm to reach a satisfactory results. Following the scalings reported in \cite{harrigan2021QAOA_non_planar,stilck2021limitations}, we will assume that $10$ layers are enough to reach the solution, and that every QAOA layer, after the routing, needs on average $\sim \sqrt{7n}$ 2-qubit gates in 2D and $3n$ 2-qubit gates in 1D \footnote{It should be noted that a noiseless SWAP gate cannot propagate an error from one qubit to another, since it will just swap the error. Therefore, a more sophisticated model could account for this by distinguishing between the SWAP gates and other 2-qubit gates, as well as adding other sources and types of noise. We still expect our scalings to hold in that case.}. In this case, a circuit depth $D_{2D} \geq 10\sqrt{7n}$ for the 2D case is needed, and $D_{1D} \geq 30n$ in 1D. This places us in the deep circuit regime, as seen in Table \ref{table:scalings} and Eqs. (\ref{eqn:heuristic_1D}) and (\ref{eqn:empirical_2D}). In this regime, the interpretation is the following: an error in any qubit will, with high likelihood, depolarize all the other qubits. Hence, even if there is just one error in the computation, the average output solution will not be much better than random guessing. Therefore, in order to have a good solution, we need to have a computation completely free of errors. This can only realistically happen if the error rate is as low as $p \sim 1/(nD)$. We perform the exact calculation and represent it in Fig. \ref{fig:noise-1D-2D}. We see that in this simple example, we would need error rates orders of magnitude below what is currently achievable. Namely of the order of $10^{-7}$ for the 1D architecture, and $10^{-6}$ for the 2D architecture. We remark that these figures are obtained under arguably conservative assumptions. A more realistic computation would have to include all the 1-qubit and 2-qubit gates, thus restricting even more the error rate.

\section{Conclusion}

We have studied a model that captures the propagation of errors in noisy quantum devices when the final state is a product state. This is the case, for example, when trying to find the solution to classical optimization problems. For this model, we show that a single error in one qubit is propagated rapidly to the rest of the qubits. This would place stringent restrictions on the error rates that are compatible with a quantum advantage. We estimate the required error rate to be $p \sim 1/\left(nD\right)$, where $n$ is the system size and $D$ is the circuit depth. As a consequence, assuming that our error model is representative of the circuits that solve the problem on real hardware, one would expect that noisy devices can only become useful for such problems when the error rates are extremely low, framing fault tolerance as the most realistic solution.

We emphasize that results are obtained by averaging different circuits, and it may be possible that for some particular instances they do not apply. However, due to the concentration result provided in appendix \ref{appendix:conc_bound} for low depth circuits, we expect the average to be representative of a typical result.

Our results suggest that there is a trade-off between error propagation and entanglement spread. If we want to take advantage of a quantum computer, the quantum circuit should be able to generate entanglement, but this will generally be associated with the propagation of errors. Equivalently, trying to avoid the propagation of errors may well result on not fully exploiting the whole quantum computer. Perhaps, this deserves a more careful analysis.

Finally, we note that there are other situations where error propagation may not impose the stringent conditions obtained here. For instance, the adiabatic algorithm \cite{farhi2000adiabatic,albash2018_adiabatic_qc} (and its variational extension \cite{Schiffer2021_variational_adiabatic}) is a special kind of circuit where one is always close to the ground state of a particular Hamiltonian. In fact, in this setup there are indications \cite{Schiffer2021_variational_adiabatic} that the propagation of errors is relatively mild. Additionally, in the development of quantum algorithms for quantum problems, like quantum simulation, conservation laws might also prevent the propagation of errors and thus circumvent the restrictions found in the present work. The analysis in this work and these considerations indicate that the propagation of errors should be taken into account in the design of quantum algorithms for NISQ devices.

\begin{acknowledgements}
The research is part of the Munich Quantum Valley, which is supported by the Bavarian state government with funds from the Hightech Agenda Bayern Plus. Funded by the Deutsche Forschungsgemeinschaft (DFG, German Research Foundation) – Project number 414325145 in the framework of the Austrian Science Fund (FWF): SFB F7104. RT acknowledges the Max Planck Harvard Research Center for Quantum Optics (MPHQ) postdoctoral fellowship.
\end{acknowledgements}

\bibliographystyle{unsrt}
\bibliography{references.bib}

\onecolumngrid
\appendix

\section{Proof of energy bound}
\label{appendix:energy}

\edit{
In this section we outline a proof for the energy bound given in Eq. (\ref{eqn:energy_bound}), stated as the following proposition:

\begin{proposition}[Energy bound]
\label{proposition:energy_bound}
    Let us consider a weighted graph $G=(V,E)$ with $n$ vertices and adjacency matrix $a_{ij} \geq 0$, and the cost function
    
    \begin{align}
        C=\frac{1}{2} \sum_{(i,j) \in E} a_{ij}\left(1-Z_iZ_j\right),
    \end{align} 
    with $Z \in \{-1,1\}^n$. The Max-Cut problem on this graph is defined as finding the maximum cut of the graph, $C_{\mathrm{max}}=\max_{Z}C$. Let us also consider the averaged quantum channel $\Phi_\text{avg}^{\mathcal{A}}$ defined in the main text. Then, for any translationally invariant architecture $\mathcal{A}$, it holds that
    
    \begin{align}
    \mathrm{tr}\left(C\Phi_{\mathrm{avg}}^{\mathcal{A}}\left(\rho\right)\right) \leq \frac{1}{2}\left(1-\frac{\left<q\right>}{n}\right)\left(2-\frac{\left<q\right>}{n}\right)C_{\mathrm{max}}+\left(1-\left(1-\frac{\left<q\right>}{n}\right)^2\right)C_{\mathrm{avg}},
\end{align}
where $\left<q\right>$ is the expectation value of the number of qubits that are depolarized  after the application of $\Phi_\textnormal{avg}^{\mathcal{A}}$, and $C_{\mathrm{avg}}=\mathrm{tr}(C/2^n)$.

\end{proposition} }
 \edit{In order to show proposition \ref{proposition:energy_bound}, we first provide a technical lemma that we will use for this bound, which provides an inequality on the joint probability of qubits at the output of the circuit to be unaffected by errors. To state this lemma, we first introduce some notation --- we consider the Markov chain introduced in section \ref{sec:markov_chain}, but for the purpose of this section we call one time-step as the application of the transition matrix of either one unitary gate (Eq.~(\ref{eq:transition_matrix_unitary})) or a depolarizing channel at a single qubit (Eq.~(\ref{eq:transition_matrix_noise})). We denote by $Q_i^t \in \{0, 1\}$ the random variable which indicates the state of the Markov chain at the $i^\text{th}$ qubit at time-step $t$. It is 0 (i.e.~the qubit experiences no noise) or $1$ (i.e.~the qubit experiences noise). Furthermore, for any subset of qubits $S$ , $Q_S^t = 0$ be the event $Q_i^t = 0 \ \forall \ i \in S$,
\begin{lemma}\label{lemma:markov_chain_joint_prob}
Let $A, B$ be two disjoint subsets of qubits, then
\[
\textnormal{Prob}(Q_{A\cup B}^t = 0) \geq \textnormal{Prob}(Q_A^t = 0) \textnormal{Prob}(Q_B^t = 0).
\]
\end{lemma}
\emph{Proof}: For ease of notation, we introduce $V^t_{A, B}$ to be
\[
V^t_{A, B} = \text{Prob}(Q_{A\cup B}^t = 0) - \textnormal{Prob}(Q_A^t = 0) \textnormal{Prob}(Q_B^t = 0).
\]
This measures the violation of the inequality that we intend to show at time step $t$ and in between the qubits in $A$ and $B$

We note that $V^t_{A, B} \geq 0$ is trivially true at $t = 0$ since all the random variables $Q_i^0$ are independent of each other. We now assume that this is true at time step $t$, and show that it is also true for $t + 1$. Suppose first that from $t$ to $t + 1$, we apply the transition matrix corresponding to two-qubit unitary on qubits $i, j$. Then, for any set $S$ such that $i, j \notin S$, we obtain by an application of the transition matrix in Eq.~(\ref{eq:transition_matrix_unitary})
\begin{align*}
    &\text{Prob}(Q_{S\cup \{i, j\}}^{t + 1} = 0) = \text{Prob}(Q_i^{t + 1} = 0, Q_j^{t + 1} = 0, Q_S^{t} = 0) \nonumber \\
    &\qquad = \text{Prob}(Q_{S\cup\{i, j\}}^{t} = 0) + \frac{1}{5}\sum_{q \in \{0, 1\}} \text{Prob}(Q_i^{t} = q, Q_j^{t} = 1 - q, Q_S^{t} = 0).
\end{align*}
Using that $\text{Prob}(Q_i^t = 0, Q_j^t = 1, Q_S^t = 0) = \text{Prob}(Q_{S\cup\{i\}}^t = 0) - \text{Prob}(Q_{S\cup\{i, j\}}^t = 0)$, we obtain 
\begin{align}\label{eq:update_rule}
    \text{Prob}(Q_{S\cup\{i, j\}}^{t + 1} = 0) = \frac{3}{5}\text{Prob}(Q_{S\cup\{i,j\}}^t = 0) + \frac{1}{5}\sum_{k \in \{i, j\}} \text{Prob}(Q_{S\cup\{k\}}^{t} = 0).
\end{align}
We will repeatedly use this update rule in the following analysis. There are three cases that we consider.  \\ 

\noindent \emph{Case }1: The two-qubit unitary is applied between a qubit in $A$ and a qubit in $B$. Let us denote by $a \in A$ and $b \in B$ the two qubits between which the unitary is applied, and by $A' = A \setminus \{a\}$ and $B' = B \setminus \{b\}$. From Eq.~(\ref{eq:update_rule}), it then follows that
\begin{align*}
    &\text{Prob}(Q_{A\cup B}^{t + 1} = 0) = \frac{3}{5}\text{Prob}(Q_{A \cup B}^t = 0) + \frac{1}{5}\sum_{k \in \{a, b\}}\text{Prob}(Q_{A' \cup B' \cup \{k\}}^t = 0).
\end{align*}
Furthermore, observe from the two-qubit unitary transition matrix in Eq.~(\ref{eq:transition_matrix_unitary}) that $Q_a^t = 0 \implies Q_b^t = 0$ (since both the qubits either simultaneously experience error or not), and therefore
\begin{align*}
    &\text{Prob}(Q_A^{t+1} = 0) = \text{Prob}(Q_{A \cup \{b\}}^{t + 11} = 0) = \frac{3}{5}\text{Prob}(Q_{A \cup \{b\}}^{t} = 0) + \frac{1}{5}\sum_{k \in \{a, b\}}\text{Prob}(Q_{A' \cup \{k\}}^t = 0).
\end{align*}
Similarly, it holds that
\begin{align*}
    \text{Prob}(Q_B^{t+1} = 0) = \frac{3}{5}\text{Prob}(Q_{B\cup\{a\}}^{t} = 0) + \frac{1}{5}\sum_{k \in \{a, b\}}\text{Prob}(Q_{B' \cup \{k\}}^t = 0).
\end{align*}
Note that since $\text{Prob}(Q_{B\cup\{a\}}^t = 0), \text{Prob}(Q_{B' \cup \{a\}}^t = 0), \text{Prob}(Q_{B' \cup \{b\}}^t = 0) \leq \text{Prob}(Q_{B'}^t = 0)$ we obtain that $\text{Prob}(Q_B^{t + 1} = 0) \leq \text{Prob}(Q_{B'}^{t + 1} = 0)$. Therefore, 
\begin{align*}
&V^{t + 1}_{A, B}\geq \text{Prob}(Q_A^{t + 1} = 0, Q_B^{t + 1} = 0) - \text{Prob}(Q_A^{t + 1} = 0) \text{Prob}(Q_{B'}^{t} = 0) = \frac{3}{5} V^{t}_{A' \cup \{a, b\}, B'} + \frac{1}{5}\bigg(V^t_{A' \cup \{a\}, B'} +  V^t_{A' \cup \{b\}, B'}\bigg),
\end{align*}
from which it follows that $V^{t + 1}_{A, B} \geq 0$. \\  \ \\
\noindent \emph{Case }2: The two-qubit unitary is applied on two qubits, $a_1, a_2 \in A$. Denote by $A' = A \setminus \{a_1, a_2\}$. Then, it follows from Eq.~(\ref{eq:update_rule})
\begin{align*}
    \text{Prob}(Q_{A\cup B}^{t + 1} = 0) =  \frac{3}{5}\text{Prob}(Q_{A\cup B}^{t} = 0) + \frac{1}{5}\sum_{k \in \{a_1, a_2\}}\text{Prob}(Q_{A'\cup B \cup \{k\}}^t = 0).
\end{align*}
Furthermore,
\begin{align*}
    \text{Prob}(Q_{A}^{t + 1} = 0) =  \frac{3}{5}\text{Prob}(Q_{A}^{t} = 0) + \frac{1}{5}\sum_{k \in \{a_1, a_2\}}\text{Prob}(Q_{A' \cup \{k\}}^t = 0),
\end{align*}
and $\text{Prob}(Q_B^{t + 1} = 0) = \text{Prob}(Q_B^{t} = 0)$. Therefore,
\[
V^{t + 1}_{A, B} = \frac{3}{5}V^t_{A, B} + \frac{1}{5}\bigg(V^{t}_{A' \cup \{a_1\}, B} +V^{t}_{A' \cup\{a_2\}, B}\bigg),
\]
from which it again follows that $V^{t + 1}_{A, B} \geq 0$. \\ \ \\
\noindent\emph{Case} 3: The two qubit unitary is applied on a qubit $a \in A$, and a qubit $c$ which is neither in $A$ nor in $B$. In this case, if $C = A \cup \{c\}$, we simply note that
\[
\text{Prob}(Q_{A\cup B}^{t + 1} = 0) = \text{Prob}(Q_{C\cup B}^{t + 1} = 0) \text{ and } \text{Prob}(Q_{A}^{t + 1} = 0) = \text{Prob}(Q_{C}^{t + 1} = 0)
\]
since $Q_a^{t + 1} = 0 \implies Q_c^{t + 1} = 0$. $V^{t + 1}_{A, B} \geq 0$ then simply follows from the analysis in case 2 with the qubit subsets $C$ and $B$. \\ \ \\
\noindent\emph{Case} 4: The two qubit unitary is applied on a qubits that are neither in $A$ nor in $B$. In this case, it trivially follows that $V^{t + 1}_{A, B} = V^t_{A, B} \geq 0$. 

Next, we consider the case when the transition matrix corresponding to the depolarizing channel (Eq.~(\ref{eq:transition_matrix_noise})) is applied to a qubit $i$. Note that if $i \notin A \cup B$, then trivially $V^{t + 1}_{A, B} = V^t_{A, B} \geq 0$. Consider now the case that $i \in A$. From Eq.~(\ref{eq:transition_matrix_noise}), it then follows that
\[
\text{Prob}(Q^{t+ 1}_{A \cup B} = 0) = (1 - p)^2 \text{Prob}(Q^{t}_{A\cup B} = 0) \text{ and } \text{Prob}(Q^{t+ 1}_{A} = 0) = (1 - p)^2 \text{Prob}(Q^{t}_{A} = 0),
\]
and therefore, $V^{t + 1}_{A\cup B} = (1 - p)^2 V^{t}_{A\cup B} \geq 0$. Hence, we have show that if at time step $t$, the lemma statement holds, then it holds for $t + 1$ which completes the proof of this lemma. $\hfill \square$}

\edit{
We will now see how the output energy after applying the averaged quantum channel can be upper bounded as a function of $\left<q\right>/n$, $C_{\mathrm{max}}$ and $C_{\mathrm{avg}}$. Let us denote by $M \subseteq E$ the subset of edges that are cut in the solution. That is, $M$ contains the edges $(i,j)$ such that $Z_i \neq Z_j$ in the maximum cut. The maximum cut can trivially be written as

\begin{align}
    C_{\mathrm{max}}= \sum_{(i,j) \in M} a_{ij},
\end{align}

while the average value of the cut when random guessing, $ C_{\mathrm{avg}}$, is

\begin{align}
    C_{\mathrm{avg}}=\mathrm{tr}\left(C\frac{I}{2^n}\right)=\frac{1}{2} \sum_{(i,j) \in E} a_{ij}.
\end{align}

While we do not have access to the complete probability distribution, we can upper and lower bound $\text{Prob}\left(Q_{i \cup j}^t=0\right)$. The traslational invariance implies that $\text{Prob}\left(Q_i^t=0\right)=r$ is the same for all qubits, and therefore

\begin{align}
\text{Prob}\left(Q_{i \cup j}^t=0\right) = \text{Prob}\left(Q_i^t=0 \cap Q_j^t=0\right) \leq  \text{Prob}\left(Q_i^t=0\right) = r.
\end{align}

Furthermore, from Lemma \ref{lemma:markov_chain_joint_prob} it follows that 

\begin{align} \label{eqn:2_point_prob}
\text{Prob}\left(Q_{i \cup j}^t=0\right) \geq \text{Prob}\left(Q_{i}^t=0\right)\text{Prob}\left(Q_{j}^t=0\right) = r^2.
\end{align}
}

\noindent \emph{Proof of proposition 1}: On the graph, we assume that $a_{ij} \geq 0 ~\forall (i,j)$. This is only for simplicity, but this condition could be dropped.  We note that, for a given edge, the contribution to the maximum cut is $a_{ij}$ if $(i,j) \in M$, while if  $(i,j) \in E \setminus M$ then it is $0$, since it will not be cut. On the other hand, if either $i$ or $j$ are random, after averaging the contribution is always $a_{ij}/2$, since it will be cut half of the times. Then, the total cost will be:
\begin{align}
    \mathrm{tr}\left(C\Phi_{\mathrm{avg}}\left(\rho\right)\right)=\sum_{(i,j)\in M}a_{ij}\text{Prob}\left(Q_{i \cup j}^t=0\right)+\frac{1}{2}\sum_{(i,j) \in E }a_{ij}\left(1-\text{Prob}\left(Q_{i \cup j}^t=0\right)\right) 
\end{align}
Rearranging the terms this yields:
\begin{align}
    \mathrm{tr}\left(C\Phi_{\mathrm{avg}}\left(\rho\right)\right)=\frac{1}{2}\sum_{(i,j)\in M}a_{ij}\left(1+\text{Prob}\left(Q_{i \cup j}^t=0\right)\right)+\frac{1}{2}\sum_{(i,j)\in E \setminus M}a_{ij}\left(1-\text{Prob}\left(Q_{i \cup j}^t=0\right)\right)
\end{align}
We can now use  Eq.~(\ref{eqn:2_point_prob}) to obtain

\begin{align}
    \mathrm{tr}\left(C\Phi_{\mathrm{avg}}\left(\rho\right)\right) \leq \frac{1}{2}\left(1+r\right)\sum_{(i,j)\in M}a_{ij}+\frac{1}{2}\left(1-r^2\right)\sum_{(i,j) E \setminus M }a_{ij}=\frac{1}{2}\left(1+r\right)C_{\mathrm{max}}+\left(1-r^2\right)\left(C_{\mathrm{avg}}-\frac{C_{\mathrm{max}}}{2}\right)
\end{align}

Then, setting $r=1-\left<q\right>/n$ proves proposition \ref{proposition:energy_bound}.

\section{Proof of concentration bound}\label{appendix:conc_bound}

In addition to studying the average energy of the output, it is also interesting to study the variance. Here we provide a concentration bound for local architectures and shallow circuits, for unweighted graphs of bounded degree $\Delta$. We show that, in this cases, the energy of a typical circuit is close to the average energy. This is done using Azuma-Hoeffding's inequality \cite{azuma_hoeffding}.
\edit{
\begin{proposition} [Concentration bound]
Let us consider an unweighted graph $G=(V,E)$ of degree $\Delta$. We denote by $C=\mathrm{tr}(C\Phi^{\mathcal{A}}(\rho))$ the value of the cut after applying one instance of the random quantum channel $\Phi$ of depth $D$, and by $\left<C\right>=\mathrm{tr}(C\Phi_{\mathrm{avg}}^{\mathcal{A}}(\rho))$ the value of the cut after applying the averaged quantum channel of depth $D$. Then, if $\mathcal{A}$ is a local architecture of nearest neighbor gates in $k$ dimensions,

\begin{align}
    \operatorname{Prob}\left[\left|C-\left<C\right> \right| \geq \alpha |E|\right] \leq  e^{-O\left(\frac{\alpha^{2}|E|}{\Delta^2 D^{2k}}\right)}.
\end{align}
\end{proposition} }

\edit{
\noindent\emph{Proof:} We begin by considering the 1D case, and will then extend the proof to higher dimensions.} We are considering a cost function that has $|E|$ terms, where $|E|$ is the number of edges of the graph. We can therefore consider $|E|$ different random variables, $X_i=\mathrm{tr}(\rho C_i)$. \edit{Here $X_i$ denotes the contribution of edge $i$ to the value of the cut}. The total cost energy will then be a random variable given by $C=\sum_i X_i$. We can now define $Z_t=\mathbb{E}[\sum_i X_i | X_1,...,X_t]$. That is, $Z_t$ updates the expectation value of the energy after learning the value of $t$ edges. In order to apply Azuma-Hoeffding's bound, we need to bound the quantity $|Z_i-Z_{i-1}|$, which determines how much the expectation value can change when we learn the energy of one edge. Since we are working with the 1D local model, two qubits can only be correlated if they are closer than $2D$, where $D$ is the depth. This holds if $D<n$. Since the degree is bounded by $\Delta$, learning the energy of one edge could at most update the value of $4D$ qubits, or $2D\Delta$ edges. We can then bound $|Z_i-Z_{i-1}| \leq \Delta D$. Applying Azuma-Hoeffding's inequality yields:

\begin{align}
    \operatorname{Prob}\left[\left|C-\left<C\right>\right| \geq \lambda\right] \leq 2 e^{-\frac{\lambda^{2}}{2|E|\Delta^2D^2}}.
\end{align}
If we set $\lambda$ to be a fraction of the total number of edges, $\lambda=\alpha |E|$, with $\alpha \in \left(0,1\right)$, this yields:
\begin{align}
    \operatorname{Prob}\left[\left|C-\left<C\right>\right| \geq \alpha |E|\right] \leq 2 e^{-\frac{\alpha^{2}|E|}{2\Delta^2D^2}}.
\end{align}
The same reasoning can be applied to the 2D case. In 2D, assuming that $D < O(\sqrt{n})$, two qubits can only be correlated if they are closer than $O(D)$, where $D$ is the depth. In this case, learning the energy of one edge could at most update the value of $O(D^2)$ qubits, or $O(D^2 \Delta)$ edges. We can then bound $|Z_i-Z_{i-1}| \leq O(\Delta D^2)$, and we obtain
\begin{align}
    \operatorname{Pr}\left[\left|C-\left<C\right>\right| \geq \alpha |E|\right] \leq 2 e^{-O\left(\frac{\alpha^{2}|E|}{\Delta^2D^4}\right)}.
\end{align}
The same argument holds generally for $k$ dimensions, completing the proof. \hfill $\square$

\color{black}

%\section{Alternative energy bound}

%Assuming $a_{ij} \geq 0$

%\begin{align}
%    \mathrm{tr}\left(C\Phi_{\mathrm{avg}}\left(\rho\right)\right)=\sum_{(i,j)\in M}a_{ij}P\left(Q_i \cap Q_j\right)+\frac{1}{2}\sum_{(i,j) }a_{ij}\left(1-P\left(Q_i \cap Q_j\right)\right) \\
%    \leq \sum_{(i,j)}a_{ij}P\left(Q_i \cap Q_j\right)+\frac{1}{2}\sum_{(i,j) }a_{ij}\left(1-P\left(Q_i \cap Q_j\right)\right) \\
%    \leq \frac{1}{2}\sum_{(i,j) }a_{ij}\left(1+P\left(Q_i \cap Q_j\right)\right) 
%    \leq \frac{1}{2}\left(1+\frac{\left<q\right>}{n}\right)\sum_{(i,j)}a_{ij}
%\end{align}

%This overestimates the cut when the graph is not bipartite, since it assumes that in the maximum cut all edges are cut. Allowing for negative $a_{ij}$ this would be:

%\begin{align}
%    \mathrm{tr}\left(C\Phi_{\mathrm{avg}}\left(\rho\right)\right) \leq \frac{1}{2}\left(1+\frac{\left<q\right>}{n}\right)\sum_{(i,j)}\left|a_{ij}\right|
%\end{align}
%But again, this would seem to overestimate the cut.

\section{Heuristic formula}
\label{appendix:Heuristic_formula}

\edit{Here we will derive the heuristic formula that provides the average number of depolarized qubits in the 1D case}, in Eq. (\ref{eqn:heuristic_1D}) from the main text:
\edit{
\begin{conjecture} [Heuristic 1D formula]
Let us consider the averaged quantum channel with the 1D architecture as defined in the main text, $\Phi_{\mathrm{avg}}^{\mathrm{1D}}$. Then, the expected number of depolarized qubits after applying $\Phi_{\mathrm{avg}}^{\mathrm{1D}}$ is approximately given by

\begin{align}
    \frac{\left<q\right>_{1D}}{n} \simeq \begin{cases} 
    1-(1-2p)^{\frac{9}{80} D^2} \text{ if }  D \leq \frac{5}{3}n\\
    1-(1-2p)^{\frac{3}{8}nD - \frac{5}{16}n^2} \text{ if }  D>\frac{5}{3}n.
    \end{cases}
\end{align}
\end{conjecture}
}
\edit{We will now outline the derivation of the formula.} The goal is to compute the expected number of depolarized qubits at the end of the computation, denoted as $\left<q\right>$, using the Markov chain defined in section \ref{sec:markov_chain}. The transition matrix $M$ of the Markov chain is given by $M=\prod_{t=1}^{D_M}M_{t}$, where $M_{t}$ is defined as the product of two matrices $M_t = M^\textnormal{noise} M_t^U$. 

%As explained in the definition, $D_M$ is the number of times that we apply the Markov chain, and it takes the value $D_M=D/2$, where $D$ is the circuit depth of the associated quantum channel. The state space of the Markov chain comprises of strings of zeroes and ones, $\{0,1\}^n$. Mapping this to the outcome of the quantum channel $\Phi_{\mathrm{avg}}$, the zeroes in the string determine which qubits are in the correct state, while the ones determine which qubits are depolarized. We will denote the average number of ones at a given time step $t$ as $\left<q\left(t\right) \right>$.

In every step there are therefore two matrices: $M_{\mathrm{noise}}$, which applies the noise with probability $p_M=2p-p^2$, and $M_U$, which propagates it. Since we are in 1D, we are applying layers of unitaries to all the even (or odd) pairs, and $M_U$ will just consist in applying the matrix \edit{that propagates the noise (Eq.~(\ref{eq:transition_matrix_unitary}))} to all even (or odd) pairs.

In order to see how this Markov chain behaves, we can consider the case where only one depolarizing error occurs in the computation. In this case, a bit will be flipped from zero to one at a certain time, and we will just apply $M_U$ from then on. Since $M_U$ can only propagate this error to neighbouring bits, in this case we can only have a string of ones surrounded by zeroes. As a consequence, we can build a Markov chain that counts the number of ones in the state. We call this the ones chain. \edit{Using the expression of $M_U$,} it is easy to see that it is a one-dimensional  lazy random walk on the line. One can note that this consists in applying the matrix that propagates the noise (Eq.~(\ref{eq:transition_matrix_unitary})) to both edges of the string of ones.

\begin{lemma}
    The ones chain has transition matrix $P$ on state space $\{0,1,...,n\}$. $P$ is given by
    \begin{align}
        P\left(x \to x'\right)=\begin{cases}
        \left(\frac{4}{5}\right)^2  \emph{ if } x' = x + 2\\
        \left(\frac{1}{5}\right)^2  \emph{ if } x' = x - 2\\
        2\left(\frac{4}{5}\right)\left(\frac{1}{5}\right)  \emph{ if } x' = x\\
        0 \emph{ else, }
        \end{cases}
    \end{align}
    for $ x \in \left(2,n-2\right)$ and $P(0 \to 0)=1$, $P(n \to n)=1$, $P(1 \to 0)=1/5$, $P(1 \to 2)=4/5$, $P(n-1 \to n)=4/5$, $P(n-1 \to n - 2)=1/5$.
    
\end{lemma}

We can bound the probability that the walker will be absorbed by the barrier in $0$, which will be useful later:

\begin{lemma}[Probability of absorption]
\label{lemma:collision_probability}
    The probability of reaching $0$ when starting in $1$ in the ones chain is upper bounded by $1/4$. That is, denoting $X_t$ the state after $t$ applications of the Markov chain:
    \begin{align}
        \textnormal{Prob}\left(X_t=0 | X_0=1\right) \leq \frac{1}{4}, \forall t.
    \end{align}
    
\end{lemma}
\emph{Proof:}
This can be shown using the standard techniques for random walks with absorbing barriers \cite{lawler2018introduction_stochastic}. The random walk starts in the state $1$. In the first step it goes to $0$ with probability $1/5$, and to $2$ with probability $4/5$. From that moment on, it can only reach even numbers, so we can discard all the states consisting of odd numbers. We can also rearrange the absorbing states ($0$ and $n$), so that they are first in the transition matrix. The transition matrix is then of the form
\begin{align*}
    \left(\begin{array}{c|c}
    I & S \\
    \hline
    0 & Q
    \end{array}\right).
\end{align*}

After infinite time, the probabilities that the walker is in $0$ or in $n$ will be given by $S\left(I-Q\right)^{-1}$. $Q$ is a tridiagonal Toeplitz matrix, and using standard techniques yields that the probability of reaching $0$, starting from $2$, with infinite time, is 
\begin{align}
    \lim_{t \to \infty}\text{Prob} \left(X_t=0 |X_0=2\right)=\left(1/5\right)^2\left[\frac{(16/25)^n-(1/25)^n}{(16/25)^{n+1}-(1/25)^{n+1}}\right] \leq \frac{1}{16}.
\end{align}

Then:
\begin{align}
    \lim_{t \to \infty}\text{Prob} \left(X_t=0 |X_0=1\right) \leq \frac{1}{5}+\frac{4}{5}\frac{1}{16}=\frac{1}{4}.
\end{align}
\hfill$\square$

We now have the ingredients to build the formula. The ones chain can be solved analytically using the techniques above. However, if we ignore the edges the relation $\left<q\left(t+1\right)\right>=\left<q\left(t\right) \right>+ 6/5$ holds. Hence, we can use the very simple formula $\left<q\left(t\right) \right> \simeq 3/4 \min \left(6/5t,n\right)$ instead. This is not exact, but it does not deviate much from the exact result. The $3/4$ factor accounts for the fact that, up to $1/4$ of the times, the walker can be absorbed by the barrier at $0$, in which case it stays there. We are neglecting the interaction with the absorbing barrier at $n$. Therefore, we see that an error propagates, on average, forming a cone. The area of this cone can be computed as

\begin{align}
\label{eqn:cone_area}
    A=\int_0^{D_M}\frac{3}{4} \min \left(6/5t,n\right)=H\left(D_M-\frac{5}{6}n\right)\left(\frac{3}{4}nD_M-\frac{5}{16}n^2\right)+H\left(-D_M+\frac{5}{6}n\right)\left(\frac{9}{20}D_M^2\right),
\end{align}
where $H(x)$ is the Heaviside step function.

Knowing this, we can see what happens approximately when there are possibly many errors during the computation. We do this by constructing a deterministic model as follows:
\begin{lemma}
    Let us construct a deterministic model of the propagation of errors as follows: if an error occurs in bit $j$ at time $t^*$, the value of bit $x_k$ at time $t$ is given by:
    \begin{align}
        x_k=\begin{cases}
            1 \emph{ if }  |j-k|<3/4 \min \left(6/5\left(t-t^*\right),n\right) \\
            0 \emph{ if }  |j-k|>3/4 \min \left(6/5\left(t-t^*\right),n\right).
        \end{cases}
    \end{align}
    
    Then, the average number of ones in this model is given by  $\left<q\left(t\right) \right>/n = 1 - \left(1-p_M\right)^A$.
    
\end{lemma}

In the model above every time there is an error it will propagate as a cone, whose area is given by Eq. (\ref{eqn:cone_area}). For the cases where there is only one error, this model yields the same value for the average number of ones as the Markov chain. This is by definition, since that single error would propagate as the average cone. If there are many errors, this is no longer the case, since in the Markov chain the different cones are not independent, while in the deterministic model they do not interact and are completely independent. However, this effect is very small. We therefore assume that this model provides a good approximation of the average number of ones in the Markov chain. Setting $p_M=2p-p^2 \simeq 2p$ and $D_M=D/2$ gives the final expression:

\begin{align}
    \left<q\right>_{1D}/n \simeq \begin{cases} 
    1-(1-2p)^{\frac{9}{80} D^2} \emph{ if }  D \leq \frac{5}{3}n\\
    1-(1-2p)^{\frac{3}{8}nD - \frac{5}{16}n^2} \emph{ if }  D>\frac{5}{3}n
    \end{cases}
\end{align}

\begin{figure*}
    \centering
    \includegraphics[scale=0.6]{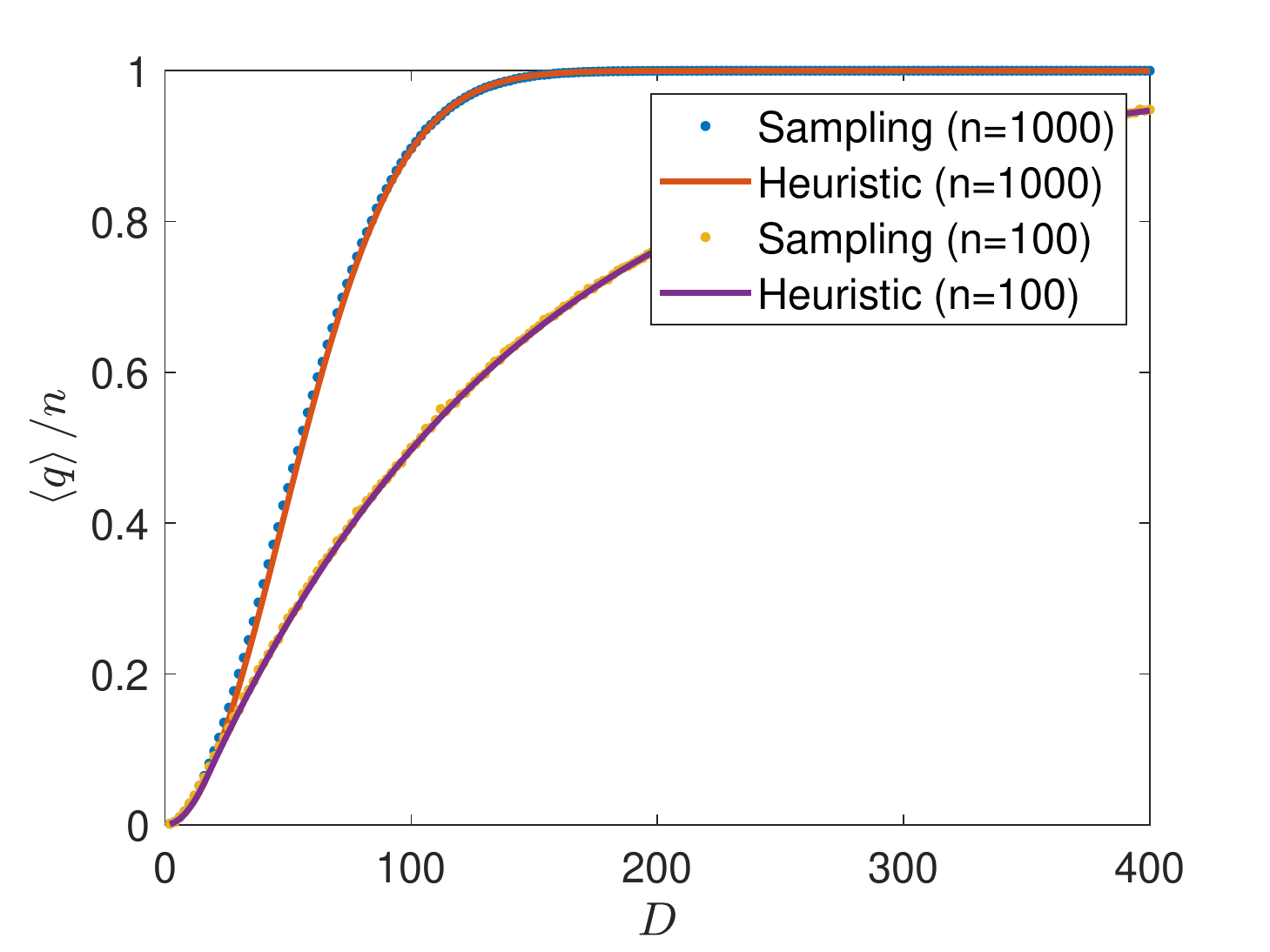}
    \caption{The heuristic formula in Eq. (\ref{eqn:heuristic_1D}) is plotted along with the results from sampling from the Markov chain for the 1D architecture. This is done for an error rate $p=10^{-3}$ and with two different system sizes, $n=100$ and $n=1000$. The heuristic formula shows good agreement with the result from sampling from the Markov chain. The number of samples taken is $50000$, and the error is upper bounded by 0.4 \%.}
    \label{fig:heuristic}
\end{figure*}

We represent this formula in Fig.~\ref{fig:heuristic}, and verify that it shows very good agreement with the result from sampling directly from the Markov chain. 

\section{Rigorous 1D bound}\label{appendix:1-D_bound}

In this section we provide a rigorous bound for $\left<q\right>/n$ in the 1D case for circuits with depth $D<n$. While this bound is not tight, it shows that the scaling in Table \ref{table:scalings} is the correct one.

\edit{
\begin{proposition}[Rigorous 1D bound] \label{proposition:rigorous_1D}
Let us consider the averaged quantum channel with the 1D architecture as defined in the main text, $\Phi_{\mathrm{avg}}^{\mathrm{1D}}$. Then, the expected number of depolarized qubits after applying $\Phi_{\mathrm{avg}}^{\mathrm{1D}}$ is lower bounded by

\begin{align}
    \left<q\right>_{1D}/n \geq c\left(1-\left(1-p\right)^{O(D^2)}\right),
\end{align}
for some $c < 3/4$.
\end{proposition}
}

\edit{To prove this proposition, we first introduce some notation. We are considering the Markov chain introduced in section \ref{sec:markov_chain}.} In this picture, an error in qubit $k$ corresponds to flipping bit $k$ from $0$ to $1$.
We define the set of all possible errors by $\mathcal{S}=\{1,..,n\}\times \{1,...,D_M\}$. Then, an error is given by the 2-tuple $(a,b) \in \mathcal{S}$. In this notation $a$ specifies the bit where the error occurred, and $b$ specifies the time step. We define an instance of errors by a subset $s \subset \mathcal{S}$ that contains all the errors that have occurred in a given run of the Markov chain. We denote by $\text{Prob}\left(Q_j^t=1\right)$ the probability that bit $j$ is in $1$ at the end of the computation. This is given by

\begin{align}
    \text{Prob}\left(Q_j^t=1\right)=\sum_{s \subset \mathcal{S}} \text{Prob}\left(Q_j^t=1\right |s)\text{Prob}\left(s\right).
\end{align}

We note that, with this Markov chain, for every instance of the errors $s \subset \mathcal{S}$ it holds that $\text{Prob}\left(Q_j^t=1\right |s) \geq \text{Prob}\left(Q_j^t=1\right | (a,b) \in s)$. That is, given an instance of errors $s$, the probability of bit $j$ being in $1$ can only decrease if we pick only one of the errors in $s$.

Let us now assume that there exists a subset of errors $A \subset \mathcal{S}$ such that, $\text{Prob}\left(Q_j^t=1\right | (a,b) \in A ) \geq c$, where $c$ is a constant. That is, $A$ contains errors such that, if one of the errors in $A$ occurs, it is enough to certify that the probability of bit $j$ being in 1 is equal or greater than $c$.
Then, we can compute 

\begin{multline}
\label{eqn:scaling_1D}
    \sum_s \text{Prob}\left(Q_j^t=1\right |s)\text{Prob}\left(s\right) \geq \sum_s \text{Prob}\left(Q_j^t=1\right |(a,b) \in s)\text{Prob}\left(s\right) \geq 
     \sum_s \text{Prob}\left(s \cap A \neq \emptyset \right)c = \\ c\left(\sum_s 1-\text{Prob}\left(s \cap A = \emptyset \right)\right)=c\left[1-\left(1-p_M\right)^{|A|}\right].
\end{multline}

We would now like to identify a subset of errors $A$ that fulfills this property. To do this, we analyze the behavior of the Markov chain. We would like to compute $\text{Prob}\left(Q_j^t=1\right | (k,t^*)) $. As explained in appendix \ref{appendix:Heuristic_formula}, if there is only one error it will propagate forming a string of ones. To compute the probability we can track the movement of the endpoints of this string. We define several random variables for this purpose:

\begin{definition}[Random walks]
\label{definition:random_walks}
    Let us consider the following random variables
    \begin{align}
        X_A(t)=X_A^0+\sum_{i=1}^{t^*}X_A^{(i)},
    \end{align}
    \begin{align}
        X_B(t)=X_B^0+\sum_{i=1}^{t^*}X_B^{(i)},
    \end{align}
    where $X_A^0=k$, and 
    \begin{align}
        X_B^0=\begin{cases}
            k+1 \emph{ with probability } 4/5 \\
            k-1 \emph{ with probability } 1/5,
        \end{cases}
    \end{align}
    
    \begin{align}
        X_A^{(i)}=\begin{cases}
            0 \emph{ if } X_A(t-1)>X_B(t-1) \\
            1 \emph{ with probability } 1/5 \emph{ if } X_A(t-1)<X_B(t-1) \\
            -1 \emph{ with probability } 4/5 \emph{ if } X_A(t-1)<X_B(t-1),
        \end{cases}
    \end{align}
    
    \begin{align}
        X_B^{(i)}=\begin{cases}
            0 \emph{ if } X_A(t-1)>X_B(t-1) \\
            -1 \emph{ with probability } 1/5 \emph{ if } X_A(t-1)<X_B(t-1) \\
            1 \emph{ with probability } 4/5 \emph{ if } X_A(t-1)<X_B(t-1). \\
        \end{cases}
    \end{align}
    
    We also define two independent random walkers as
    \begin{align}
        X_A^{\mathrm{ind}}(t)=X_A^0+\sum_{i=1}^{t^*}X_A^{\mathrm{ind},(i)},
    \end{align}
    
    \begin{align}
        X_B^{\mathrm{ind}}(t)=X_B^0+\sum_{i=1}^{t^*}X_B^{\mathrm{ind},(i)},
    \end{align}
    where
    
    \begin{align}
        X_A^{\mathrm{ind},(i)}=\begin{cases}
            1 \emph{ with probability } 1/5\\
            -1 \emph{ with probability } 4/5,
        \end{cases}
    \end{align}
    
    \begin{align}
        X_B^{\mathrm{ind},(i)}=\begin{cases}
            -1 \emph{ with probability } 1/5\\
            1 \emph{ with probability } 4/5. 
        \end{cases}
    \end{align}
\end{definition}

With the definitions above, it holds that
\begin{align}
\label{eqn:1_error_to_random_walk}
    \text{Prob}\left(Q_j^t=1\right | (k,t^*)) = \text{Prob}(X_B(t) \geq j \cap X_A(t) \leq j).
\end{align}

The random walks $X_A$ and $X_B$ defined above track the end points of the string of ones. The difficulty here is that they are not independent random walks. They behave independently until they cross: in that case they stop. However, we can bound the probability using $X_A^{\mathrm{ind}}$ and $X_B^{\mathrm{ind}}$, which are just biased random walks in the line, with every step being independent on the rest.

\begin{lemma}
\label{lemma:independent_random_walks}
Let us consider the random variables $X_A$, $X_A^{\mathrm{ind}}$, $X_B$ and $X_B^{\mathrm{ind}}$as defined in definition \ref{definition:random_walks}. Then,

    \begin{align}
        \textnormal{Prob}\left(Q_j^t=1 | (k,t^*) \right) = \textnormal{Prob}(X_B(t) \geq j \cap X_A(t) \leq j) \geq \textnormal{Prob}\left(X_B(t)^{\mathrm{ind}} \geq j\right)\textnormal{Prob}\left(X_A(t)^{\mathrm{ind}} \leq j\right) - \frac{1}{4}.
    \end{align}
\end{lemma}
\emph{Proof:} We want to compute $\text{Prob}(X_B(t) \geq j \cap X_A(t) \leq j)$. We note that the two random walks are completely independent until the moment they cross: in that case they stop. We can denote this event as $C$, and no crossing as $NC$. Assuming that $t<n/2$, from Lemma \ref{lemma:collision_probability} we know that the probability that they cross is upper bounded, $\text{Prob}(C) \leq 1/4$. Then, we can write:

\begin{align}
\label{eqn:random_walks_prob}
    \text{Prob}\left(X_A(t) \geq j \cap X_B(t) \leq j\right)=\text{Prob}\left(X_A(t) \geq j \cap X_B(t) \leq j |NC\right)\text{Prob}\left(NC\right).
    \end{align}
We can now consider the two random walks $X_A^{\mathrm{ind}}$ and $X_B^{\mathrm{ind}}$. Their behavior is the same as $X_A$ and $X_B$ provided there is no crossing. That is:
    \begin{align}
    \text{Prob}\left(X_A(t) \geq j \cap X_B(t) \leq j |NC\right)=\text{Prob}\left(X_A^{\mathrm{ind}}(t) \geq j \cap X_B^{\mathrm{ind}}(t) \leq j |NC\right).
    \end{align}

Using now the law of total probability we get:

\begin{multline}
    \text{Prob}\left(X_A^{\mathrm{ind}}(t) \geq j \cap X_B^{\mathrm{ind}}(t) \leq j |NC\right)\text{Prob}(NC) = \\ \text{Prob}\left(X_A^{\mathrm{ind}}(t) \geq j\right)\text{Prob}\left(X_B^{\mathrm{ind}}(t) \leq j\right) - \text{Prob}\left(X_A^{\mathrm{ind}}(t) \geq j \cap X_B^{\mathrm{ind}}(t) \leq j |C\right)\text{Prob}(C) \geq \\ \text{Prob}\left(X_A^{\mathrm{ind}}(t) \geq j\right)\text{Prob}\left(X_B^{\mathrm{ind}}(t) \leq j\right) -\frac{1}{4}.
\end{multline}

The result then follows immediately from Eq.  (\ref{eqn:1_error_to_random_walk}).\hfill$\square$

We know now how to bound the probability $\text{Prob}\left(Q_j^t=1\right | (k,t^*))$ using two biased random walks on the line. This allows us to obtain the \edit{combinations of $(k,t^*)$ that we are interested in}:
\begin{lemma}
Let us consider an error $(k,t^*)$ such that
\begin{align}
    \left|j-k\right| \leq \frac{3}{5}\left(t-t^*\right)-\sqrt{2\left(t-t^*\right) \ln \frac{1}{1-\sqrt{\frac{1}{4}+c}}   }.
\end{align}
Then, $\textnormal{Prob}\left(Q_j^t=1\right | (k,t^*)) \geq c$.
\end{lemma}
\edit{
\noindent\emph{Proof:} Since $X_A^{\mathrm{ind}}$ and $X_B^{\mathrm{ind}}$ are a sum of independent random variables, we can bound the probability that they deviate from the mean using Hoeffding's inequality \cite{1963_hoeffding_inequality}:
\begin{align}
    \text{Prob}\left(X_A^{\mathrm{ind}}(t)-E\left[X_A^{\mathrm{ind}}(t)\right] \geq \alpha \right) \leq \exp \left(-\frac{\alpha^2}{2\left(t-t^*\right)}\right).
\end{align}
The expectation values can easily be computed as $E\left[X_A^{\mathrm{ind}}(t)\right]=k-3/5(t-t^*)$ and $E\left[X_B^{\mathrm{ind}}(t)\right]=k+3/5(1+t-t^*)$. Let us denote $d_A=j-E\left[X_A^{\mathrm{ind}}(t)\right]$. Then,
\begin{align}
    \text{Prob}\left(X_A^{\mathrm{ind}}(t) \leq j\right) = 1 - \text{Prob}\left(X_A^{\mathrm{ind}}(t) \geq j\right) = 1 - \text{Prob}\left(X_A^{\mathrm{ind}}(t) -E\left[X_A^{\mathrm{ind}}(t)\right] \geq j -E\left[X_A^{\mathrm{ind}}(t)\right]\right) \geq 1-\exp \left(-\frac{d_{A}^2}{2\left(t-t^*\right)}\right),
\end{align}
Let us now assume, without loss of generality, that $j<k$. Then,
\begin{align}
    \text{Prob}\left(X_B^{\mathrm{ind}}(t) \geq j\right)  \geq \text{Prob}\left(X_A^{\mathrm{ind}}(t) \leq j\right).
\end{align}
Therefore, we obtain that
\begin{align}
    \text{Prob}\left(Q_j^t=1\right | (k,t^*)) \geq \left( 1-\exp \left(-\frac{d_{A}^2}{2\left(t-t^*\right)}\right)  \right)^2 -\frac{1}{4} \geq c.
\end{align}

Solving for $d_A$ then yields

\begin{align}
    d_A \geq \sqrt{2(t-t^*)\ln \frac{1}{1-\sqrt{\frac{1}{4}+c}}}
\end{align}
And therefore:
\begin{align}
    \left|k-j\right| \leq \frac{3}{5}(t-t^*) - \sqrt{2\left(t-t^*\right)\ln\frac{1}{1-\sqrt{\frac{1}{4}+c}}}.
\end{align}
For symmetry reasons, this calculation holds as well when $j>k$.} \hfill $\square$

We have therefore found a region such that $\text{Prob}\left(Q_j^t=1\right | (k,t^*))\geq c$. Integration over $t^*$ will yield the area of such region:

\begin{align}
|A|=\frac{3}{5}t^2-\frac{4}{3}\sqrt{2\ln \frac{1}{3/4-c}}t^{3/2}.
\end{align}

Then, using Eq. (\ref{eqn:scaling_1D}) we can bound
\begin{align}
    \left<q\right>/n \geq c\left(1-\left(1-p_M\right)^{\frac{3}{5}t^2-\frac{4}{3}\sqrt{2\ln \frac{1}{1-\sqrt{\frac{1}{4}+c}}}   t^{3/2}   }\right). 
\end{align}
\edit{
Then, setting $p_M=2p-p^2 \simeq 2p$ and $t=D/2$ yields the result in proposition \ref{proposition:rigorous_1D}.
}
\end{document}